\documentclass[fleqn,usenatbib]{mnras}

\usepackage{newtxtext,newtxmath}

\usepackage[T1]{fontenc}

\DeclareRobustCommand{\VAN}[3]{#2}
\let\VANthebibliography\thebibliography
\def\thebibliography{\DeclareRobustCommand{\VAN}[3]{##3}\VANthebibliography}

\usepackage{graphicx}	
\usepackage{amsmath}	
\usepackage{pdflscape}
\usepackage{float}
\usepackage[aboveskip=0pt]{caption}
\usepackage{booktabs,siunitx}

\title[Comptonization region in AGNs and its influence]{Influence of Comptonization region over the ambiance of accretion disc in active galactic nucleus}

\author[K. Sriram et al.]{
K. Sriram,$^{1}$
D. Nour,$^{1}$
and C. S. Choi$^{2}$
\\
$^{1}$Department of Astronomy, Osmania University, Hyderabad 500007, India\\
$^{2}$Korea Astronomy and Space Science Institute, Daejeon 34055, Republic of Korea
}

\date{Accepted XXX. Received YYY; in original form ZZZ}

\pubyear{2015}

\begin{document}
\label{firstpage}
\pagerange{\pageref{firstpage}--\pageref{lastpage}}
\maketitle

\begin{abstract}
Understanding the radiative and physical structures of inner region of a few 100 pc of AGNs is important to constrain the causes of their activities. Although the X-ray emission from the Comptonization region/corona and the accretion disc regulates the broad line emission regions and torus structures, the exact mutual dependency is not understood well. We performed correlation studies for X-ray, mid-infrared, and different components of Balmer emission lines for the selected sample of AGNs. Almost 10 different parameters and their inter-dependencies were explored in order to understand the underlying astrophysics. We found that the X-ray luminosity has a linear dependency on the various components of broad Balmer emission lines (e.g. L$_{\text{2-10 keV}}$ $\propto$ L$^{0.78}_{\text{H}\beta^{\text{B}}}$) and found a strong dependency on the optical continuum luminosity (L$_{\text{2-10 keV}}$ $\propto$ L$^{0.86}_{5100\,\text{\AA}}$). For a selected sample, we also observed a linear dependency between X-ray and mid-infrared luminosity (L$_{\text{2-10 keV}}$ $\propto$ L$^{0.74}_{6\,\upmu \text{m}}$). A break point was observed in our correlation studies for X-ray power-law index, $\Gamma$, and mass of black hole at $\sim$ log (M/M$_{\odot}$) = 8.95. Similarly the relations between $\Gamma$ and FWHM of H$\alpha$ and H$\beta$ broad components show breaks at FWHM$_{\text{H}\alpha}$= 7642$\pm$657 km s$^{-1}$ and FWHM$_{\text{H}\beta}$ = 7336$\pm$650 km s$^{-1}$. However, more data are required to confine the breaks locations exactly. We noted that $\Gamma$ and Eddington ratios are negatively correlated to Balmer decrements in our selected sample. We analyzed and discussed about the implications of new findings in terms of interaction AGN structures. 
\end{abstract}




\section{Introduction}

Gas and dust accretion plays a key role in the growth of super massive black holes (SMBHs) in the galaxies across their redshift range. Some of the galaxies having a brightened core due to a higher radiative emission are known as active galactic nucleus (AGNs). The primary components of nucleus comprise a central source, a broad line region (BLR), a narrow line region (NLR), and a molecular dust/grain torus. The BLR is closest to the accretion disc powering the AGN activity and dynamically Keplerian in nature with additional velocity components (Done \& Krolik 1996;  Collin et al. 2006; Osterbrock \& Ferland 2006). Fraction of AGNs exhibits broad emission lines with a full width at half maximum (FWHM) from over 1000 km s$^{-1}$ to some 15000 km s$^{-1}$(Czerny 2019), this emission originates from BLR that is probably located at a distance of 1000 R$_{g}$ from the SMBH (R$_{g}$= GM/c$^2$; the gravitational radius of the black hole; Peterson 2008) with a gas density of 10$^{9}$ -- 10$^{11}$ cm$^{-3}$ (Peterson 1993). The outer edge of BLR is surrounded by a dust/molecular torus (Netzer \& Laor 1993; Suganuma et al. 2006). The emission lines arising from NLR have a FWHM of 200 -- 1000 km s$^{-1}$ and a temperature of T $\sim$ 10$^{4}$ K with a low gas density (N$_{e}$ $\sim$ 10$^{2}$ - 10$^{6}$ cm$^{-3}$; Vaona et al. 2012).\\
In order to understand the accretion process and the structure of emitting regions/locations in AGNs, studies of broad-band spectral energy distributions (SEDs) and their temporal variations are necessary. The X-ray emission is produced in the innermost region or above the accretion disc by hot electrons i.e. corona (Ramos Almeida \& Ricci 2017), where ultraviolet (UV) photons are Compton up-scattered (Haardt \& Maraschi 1993) which results in a power-law continuum above 2 keV. The modified black body emission associated with the Keplerian portion of the accretion disc peaks around optical/UV (Sanders et al. 1989; Elvis et al. 1994). Some of the emitted UV radiation is reprocessed in a structure often known as “dusty torus”, located around tens of pc scale environment of AGNs, and is re-emitted in infrared/mid-infrared (MIR) (Suganuma et al. 2006;  Kishimoto et al. 2007; Burtscher et al. 2013; Ramos Almeida \&  Ricci 2017; Leftley et al. 2018). Recent studies based on infrared observations with Very Large Telescope Interferometer (VLTI) suggest that the distribution of dust in the torus need not to be confined in a classical thick toroidal structure but possibly originates from two components i.e. an equatorial thin disc and a hollow dusty cone toward the polar region (see Honig \& Kishimoto 2017; Stalevski et al. 2017, 2019). The hollow dusty cone is probably due to the puffed-up disc caused by the infrared radiation pressure which releases the inflowing gas from the gravitational potential of the black hole (Honig 2019). This structure is considered to be an extended polar feature probably from a dusty wind located at the edge of the narrow line region (Lopez-Gonzaga et al. 2014; Honig \& Kishimoto 2017; Lopez-Gonzaga et al. 2016; Stalevski et al. 2017;  Leftley et al. 2018). The MIR radiation is also considered to be emitted from this non-classical torus structure (Asmus et al. 2016) and the spectrum is often modeled by a warm component with a black body temperature around 300 K (e.g. Edelson \& Malkan 1986). The orientation of the dusty molecular obscuring torus plays a key role in the characterization of the distinct classes of AGNs in the unification scheme (for more detail see, Antonucci 1993; Urry \& Padovani 1995; Netzer 2015).\\
A correlation between MIR and X-ray was observed for a small sample of AGNs (Elvis et al. 1978; Glass et al. 1982). Later studies reinforced the correlation from a larger sample with different correlation strengths for type 1 and type 2 AGNs (e.g. Lutz et al. 2004; Ramos Almeida et al. 2007). Gandhi et al. (2009) studied a sample of 22 AGNs and found a linear correlation between MIR and X-ray emission L$_{12\,\upmu \text{m}}$ $\propto$ L$_{2-10\,\text{ keV}}^{1.11\pm0.07}$ for local Seyfert galaxies. The MIR and X-ray emission in AGNs are strongly correlated in type 1 and type 2 AGNs also in case of radio quite or radio loud AGNs. Therefore, this correlation exists in all AGNs irrespective of the object nature and underlying physics (Asmus et al. 2015). Asmus et al. (2015) found that luminous AGNs have relatively more X-ray emission than the prediction of the local linear relation between MIR and X-ray luminosity given by Gandhi et al. (2009). Chen et al. (2017) used a large sample of 3247 type 1 AGNs and reported a bi-linear relation between L$_{2-10\,\text{keV}}$ and L$_{6\,\upmu \text{m}}$  i.e. luminous type 1 quasars have shallower correlation than the correlation exhibited by the local Seyfert galaxies. A few other studies show that there exists a luminosity dependent relation between L$_{2-10\,\text{keV}}$ and L$_{6\,\upmu \text{m}}$ (Stern et al. 2014; Assef et al. 2016). 
Jin et al. (2012a) observed a break in the relation between FWHM of broad H$\beta$ and X-ray power-law index (2-10 keV; $\Gamma$) at FWHM = 4000 km s$^{-1}$ and noted that this break may have connection with differences among Population A and B AGNs.
They also found a break in the relation between SMBH mass (M$_{\text{BH}}$) and $\Gamma$ at $\log$ (M$_{\text{BH}}$/M$_{\odot}$) = 8, these two relations are important to understand the AGNs and their evolution. Lakicevic et al. (2018) and references therein have noticed breaks in luminosities between narrow and broad line Seyfert galaxies (NLS1s and BLS1s), while Lakicevic et al. (2022) found that these breaks more likely happen at FWHM (H$\beta$) $\approx$ 4000 km s$^{-1}$ (than 2000 km s$^{-1}$, where NLS1s and BLS1s separate). Lakicevic et al. (2022) suggested that these breaks may be a consequence of the AGN orientation (NLS1s are seen in lower inclination angles than BLS1s, see Zhang \& Wu 2002).\\
Lakicevic et al. (2017) performed MIR and optical correlation studies for 82 type 1 AGNs obtained from Spitzer and Sloan Digital Sky Survey (SDSS) DR12 databases. They found that optical and MIR starburst diagnostics are different for their sample but noticed some correlations between optical and IR spectral parameters. In this study, we explored this sample to understand the X-ray, optical, and MIR association. We independently reduced and analyzed the SDSS spectra for 53 AGNs that cover both H$\alpha$ and H$\beta$ in their spectra. We investigated not only the degree of correlation between X-ray luminosity and different components of optical emission lines (H$\alpha$ \&  H$\beta$), but also the spectral parameters of optical emission lines such as FWHM, equivalent width (EW), and Balmer decrements how they correlate with $\Gamma$ and Eddington ratio ($\lambda_{\text{Edd}}$).
We also studied the break phenomena observed in the relation between $\Gamma$ and FWHM \& M$_{\text{BH}}$ and compared those with the results by Jin et al. (2012a).
\section{Data Reduction and Analysis for SDSS Spectra} \label{sec:floats}

Using the SDSS sample given by Lakicevic et al. (2017), we obtained 72 type 1 AGN spectra where both H$\alpha$ and H$\beta$ emission lines were observed. Their optical spectra were downloaded from the SDSS data release 12 (DR12) (Alam et al. 2015). The observations were performed using the Sloan foundation 2.5-meter telescope located at Apache Point Observatory in New Mexico. 
The following method was adopted to reduce the spectra and obtain the various parameters associated with the different emission lines. 

\begin{enumerate}
 \item Galactic extinction corrections were applied to the spectra based on the parameterization by Fitzpatrick (1999) and the galactic extinction coefficients provided by Schlegel et al. (1998).

 \item Wavelengths and fluxes were corrected for redshifts.
 \item Correction of host galaxy contribution: The SDSS optical spectra contain contributions from both the central engine of AGN and host galaxy. Therefore, the elimination of host galaxy component is necessary to study the central engine properties. Vanden Berk et al. (2006) study shows that a galaxy spectrum can be considered as a linear combination of eigenspectra, hence, following this assumption, one can decompose an AGN spectrum into two parts i.e. pure-QSO (QSO) part and pure-galaxy (GAL) part using two sets of eigenspectra.\\
 
\begin{figure}
	\includegraphics[width=\columnwidth]{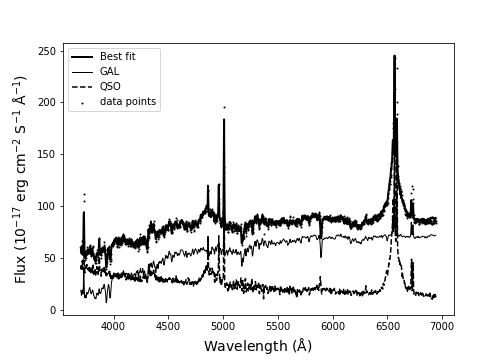}
    \caption{Example of spectral decomposition of J145901.36+611353.59 into QSO and GAL parts. The bold line shows the best-fitting model, solid line corresponds to the GAL part, and dashed line the QSO part. The dots represent the observed data points in the figure.}
    \label{fig:1}
\end{figure}

\begin{figure}

	\includegraphics[width=\columnwidth]{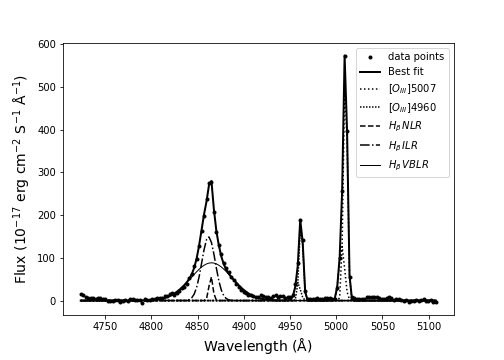}
    \caption{H$\beta$ and [OIII] emission line fits for J093701.05+010543.79 and different fitting components are shown in different markers (see the legend).}
    \label{fig:2}
\end{figure}
To perform this decomposition, we used the principal component analysis (PCA) where we adopted procedure from Lakicevic et al. (2017) by using the first 10 QSO eigenspectra given by Yip et al. (2004a) for high luminosity and low redshift (CZBIN1) and the first 5 galaxy (GAL) eigenspectra of Yip et al. (2004b). 
We first re-binned the 15 eigenspectra (10 QSO + 5 GAL) and SDSS spectra so that all they have the same wavelength range and bins. Then we fitted each spectrum with a linear combination of the 15 eigenspectra using the following relation:
\begin{equation}
f(\lambda)=\sum_{i=1}^{10} a_i Q_i(\lambda) + \sum_{j=1}^{5} a_j G_j(\lambda)
\end{equation}
where f($\lambda$) is the best-fitting flux, $a_i$ and $a_j$ are eigencoefficients, $Q_i$ and $G_j$ represent eigenspectra of QSO and GAL (host galaxy), respectively.
By subtracting absorption lines and host galaxy continuum from the observed spectra, we got the spectra of respective AGNs.

An example of the decomposition is shown in Fig. 1. 
 \item Subtraction of continuum: Using the continuum windows given by Kuraszkiewicz et al. (2002), the selected windows were interpolated and fitted using a power-law model to determine the continuum levels which were later subtracted.
 \item Subtraction of FeII lines: FeII lines contribute to the emission in the region 4000 - 5500 {\AA}. FeII contributions were subtracted using the software suggested by Kovacevic et al. (2010) and Shapovalova et al. (2012).  
\begin{figure}

	\includegraphics[width=\columnwidth]{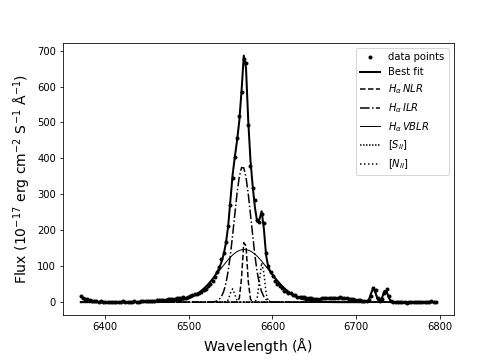}
    \caption{H$\alpha$ emission line fit for J093701.05+010543.79 and different fitting components are shown in different markers (see the legend).}
    \label{fig:3}
\end{figure} 
 
 \item Fit of emission lines: We performed fitting in two spectral regions: 4000 - 5500 {\AA} to study H$\beta$ \& [OIII] emission lines and 6300 - 8500 {\AA} to study H$\alpha$ where [NII] and [SII] lines also contribute. H$\beta$ and H$\alpha$ emission lines were fitted with the model of three Gaussian components which correspond to emission originating from narrow line region (NLR), intermediate line region (ILR), and very broad line region (VBLR). The combination of ILR and VBLR is often termed as BLR (Popovic et al. 2004). Examples of the fits are shown in Fig. 2 and Fig. 3.

Narrow emission lines i.e. [NII] \& [SII] and narrow component of emission lines H$\beta$, [OIII] 4959 {\AA}, 5007 {\AA}, and H$\alpha$ originates from the same NLR, therefore, they are considered to have same width and shift. [OIII] 4959 {\AA}, 5007 {\AA} lines are fitted with another blue-shifted component that is associated with outflows (Aoki et al. 2005). We considered the ratio between [OIII] doublet components as 1:2.99 (Dimitrijevic et al. 2007) and for [NII] the ratio is 3.

 \item After applying all the corrections to the emission lines, we calculated various parameters such as luminosity, full width at half-maximum (FWHM), and equivalent width (EW) for H$\beta$ and H$\alpha$ components. Luminosities of emission lines components were calculated using the formula given by Peebles (1993) by adopting the cosmological parameters $\Omega_M$ = 0.3, $\Omega_\Lambda$ = 0.7, and $\Omega_k$ = 0, H$_0$ = 70 km s$^{-1}$ Mpc$^ {-1}$. Similarly we calculated the continuum luminosity where the continuum flux was measured at 5100$\text{\AA}$. FWHMs were calculated from the fitted Gaussians. EWs of emission lines were measured after subtraction of the host galaxy contribution, with respect to continuum below the lines (see Kovačević et al. 2010). These values are tabulated in Table 1 and Table 2. In Table 1, we listed the derived parameters of H$\beta$ emission line components as follows. Object number; SDSS name; RA and Dec; redshift; EWs of NLR, ILR, VBLR components; FWHMs of NLR, ILR, VBLR components; luminosities of NLR, ILR, VBLR components. In  Table 2, we also listed the same derived parameters of H$\alpha$ emission line components.
\end{enumerate}

\section{Data Collection in other wavelengths} \label{sec:style}

The correlation of X-ray and optical behaviour of AGNs is important as it helps to unveil the underlying physics i.e. how X-rays from corona and accretion disc affect the optical emissions in the BLRs. We adopted X-ray luminosities in 2-10 keV energy band for 53 sources from the archival data shown in Col. (13) of Table 3. 

We adopted MIR luminosities at 6 $\upmu$m and 12 $\upmu$m calculated by Lakicevic et al. (2017) for the 53 AGNs. We used soft X-ray luminosities (0.1-2.4 keV) from Anderson et al. (2007) for 27 sources. All the parameters obtained and calculated are listed in Table 3 for the selected sample of 53 type 1 AGNs.

In Table 3, we listed the following. Object number; continuum luminosity at 5100 {\AA}; MIR luminosities L$_{6\,\upmu \text{m}}$ and L$_{12\,\upmu \text{m}}$; X-ray luminosity; X-ray photon index in 2 - 10 keV energy band; soft X-ray luminosity; FWHMs of broad components (ILR+VBLR) of H$\beta$ and H$\alpha$; mass of the black hole; Eddington ratio; broad Balmer decrement; references for L$_{2-10\, \text{keV}}$ and $\Gamma$.

\section{RESULTS AND DISCUSSIONS}
We performed correlation studies for various wavelength dependent parameters to investigate the inter-dependency of emission regions in the selected sample of AGNs. The Spearman’s correlation coefficients ($\uprho$) and probabilities (P) for each correlation between parameters are tabulated in the Table 4.
\subsection {X-ray and optical emission}
\subsubsection{Balmer lines luminosity}
\begin{figure*}
	\includegraphics[width=\textwidth]{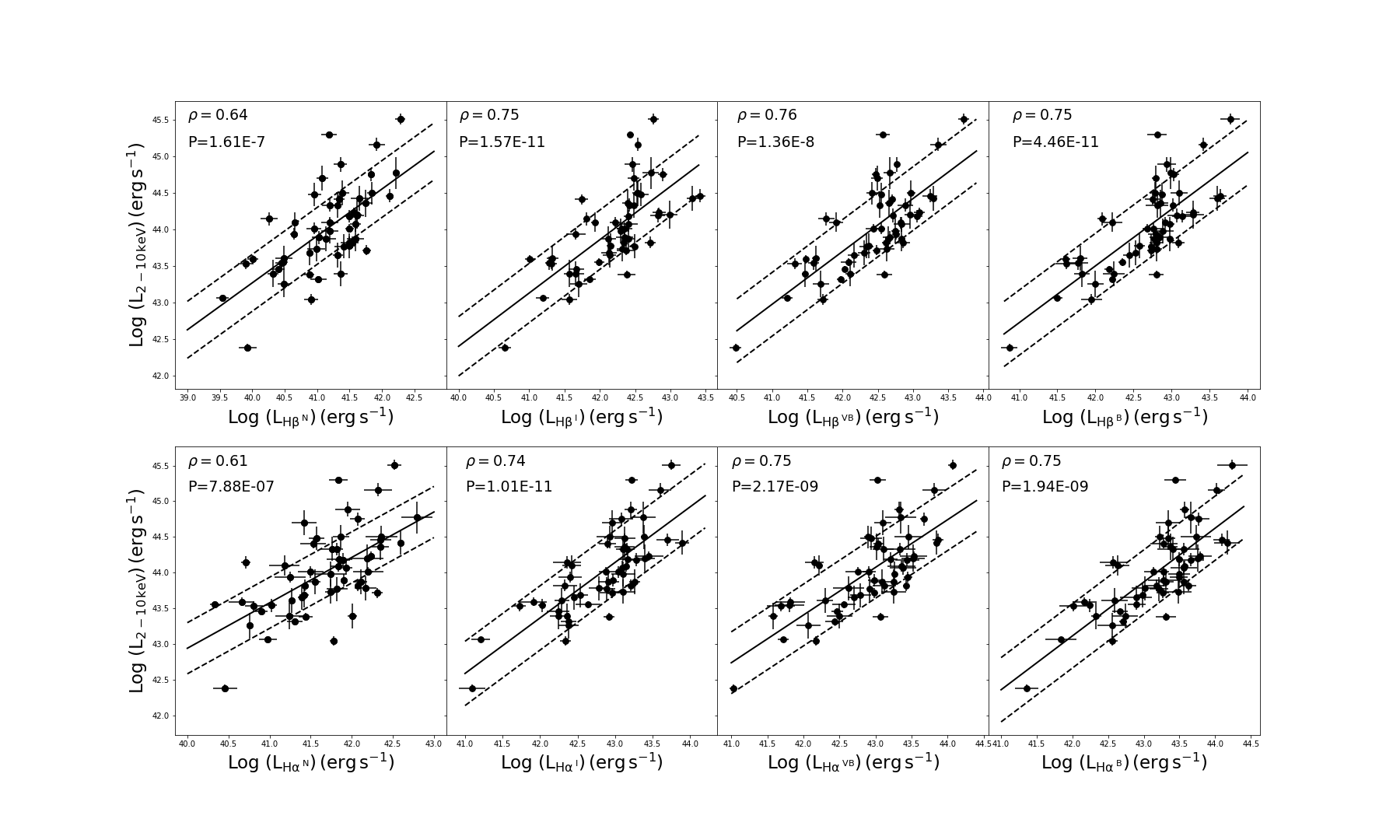}
    \caption{Correlation between luminosity of H$\beta$ and H$\alpha$ components and  L$_{2-10\,\text{keV}}$. The solid line in each panel shows the regression line along with $\pm 1 \sigma$ region represented by the dashed lines.}
    \label{fig:4}
\end{figure*}
Balmer lines (i.e. H$\alpha$ and H$\beta$) luminosities positively correlate with X-ray luminosity (2-10 keV) and different components of the line show different degree of correlation. VBLR component shows stronger correlation coefficient ($\uprho$=0.76, P=1.36E-8 for H$\beta$ and $\uprho$=0.75, P=2.17E-9 for H$\alpha$) than the coefficient ($\uprho$=0.64, P=1.61E-7 for H$\beta$ and $\uprho$=0.61, P=7.88E-7 for H$\alpha$) of NLR. This confirms that VBLR is the closest to the central engine where X-ray emission is produced via Compotonization in corona and in the accretion disc while NLR is further away. The ILR component ($\uprho$=0.75, P=1.57E-11 for H$\beta$ and $\uprho$=0.74, P=1.01E-11 for H$\alpha$) was found to be similar to VBLR and hence it is difficult to consider that it is occupying a separate location in the AGN structure. 
Correlation coefficients are shown on the upper left corner of each panel in Fig. 4 and the obtained relations for H$\beta$ components are as follows.

\begin{equation}
\log(L_{2-10\,\text{keV}})=(0.64\pm0.10) \log(L_{\text{H}\beta^{\text{N}}})+(17.59\pm4.13)
\end{equation}

\begin{equation}
\log(L_{2-10\,\text{keV}})=(0.73\pm0.10) \log(L_{\text{H}\beta^\text{I}})+(13.24\pm4.44)
\end{equation}
\begin{equation}
\log(L_{2-10\, \text{keV}})=(0.73\pm0.08) \log(L_{\text{H}\beta^{\text{VB}}})+(13.31\pm3.81)
\end{equation}
\begin{equation}
\log(L_{2-10\, \text{keV}})=(0.78\pm0.09) \log(L_{\text{H}\beta^\text{B}})+(10.91\pm3.91)
\end{equation}

We also found the luminosities of H$\alpha$ components to correlate strongly with L$_{2-10\,\text{keV}}$ and the relations are as follows.
\begin{equation}
\log(L_{2-10\,\text{keV}})=(0.64\pm0.11) \log(L_{\text{H}\alpha^\text{N}})+(17.45\pm4.76)
\end{equation}
\begin{equation}
\log(L_{2-10\,\text{keV}})=(0.77\pm0.08) \log(L_{\text{H}\alpha^\text{I}})+(10.69\pm3.83)
\end{equation}
\begin{equation}
\log(L_{2-10\,\text{keV}})=(0.74\pm0.08) \log(L_{\text{H}\alpha^\text{VB}})+(15.31\pm3.62)
\end{equation}
\begin{equation}
\log(L_{2-10\,\text{keV}})=(0.75\pm0.09) \log(L_{\text{H}\alpha^\text{B}})+(11.39\pm3.75)
\end{equation}

The obtained slopes of regression lines are close to those reported by Jin et al. (2012b), slope = 0.83$\pm$0.03, for the broad component of H$\beta$ and H$\alpha$. A similar study by Panessa et al. (2006) derived the relation between  L$_{\text{X}}$ and  L$_{\text{H}\alpha^{\text{B}}}$ and found the slope to be 0.74$\pm$0.21. These studies along with the slopes obtained by our sample indicate that X-ray is affecting the BLR  emission in different AGNs satisfying a similar relationship of L$_\text{X}$--L$_{\text{H}\beta^{\text{B}}}$ and L$_\text{X}$--L$_{\text{H}\alpha^{\text{B}}}$. As a strong correlation exists between X-ray and broad component of emission lines, studying the X-ray activity is necessary to explore the BLR.\\
Moreover, we noted that, for our sample of 53 type 1 AGNs, X-ray and optical luminosities at 5100 {\AA} are highly correlated (Fig. 5) satisfying a linear regression relation:

\begin{equation}
\log(L_{2-10\,\text{keV}})=(0.86\pm0.10) \log(L_{5100\text{\AA}})+(5.89\pm4.65)
\end{equation}
\begin{figure}
    \includegraphics[width=\columnwidth]{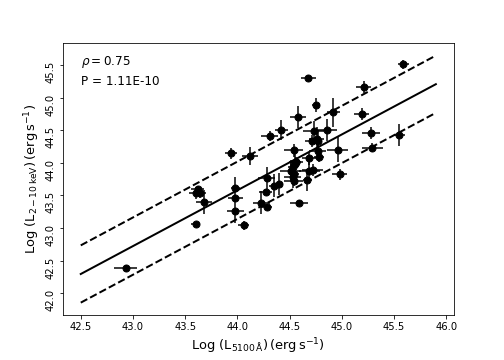}
    \caption{Correlation between L$_{2-10\,\text{keV}}$ and continuum luminosity (L$_{5100\,\text{\AA}}$).}
    \label{fig:5}
\end{figure}

\subsubsection {Eddington ratio vs FWHM, EW }
The ratio of bolometric luminosity to Eddington luminosity is known as Eddington ratio ($\lambda_{\text{Edd}}$). It is an important parameter to understand AGNs evolution because the bolometric luminosity is connected to the mass accretion rate. Studying the relation between $\lambda_{\text{Edd}}$ and optical parameters of emission lines enables us to understand how the growth of the SMBH is affecting the nearby region of AGNs. The Eddington ratios were estimated using the following relations (Elvis et al. 1994; Marconi et al. 2004):
 \begin{equation}
\lambda_{\text{Edd}} = L_{\text{bol}}/ L_{\text{Edd}}
\end{equation}
\begin{equation}
  L_{\text{bol}} = 20\times L_{2-10\,\text{keV}}\,(\text{erg}\, \text{s}^{-1})
\end{equation}
\begin{equation}
  L_{\text{Edd}}=1.26\times 10^{38} \text{M}_{\text{BH}}/\text{M}_\odot\, (\text{erg}\, \text{s}^{-1})
\end{equation} 

As shown in Fig. 6 (upper left panel), FWHM of the broad component of H$\beta$ shows a negative correlation with $\lambda_{\text{Edd}}$ ($\uprho$ = -0.75), indicating that AGNs with a low accretion rate tends to have broader emission lines compared to AGNs having a high accretion rate (Bian \& Zhao 2003 and references therein).  EW vs $\lambda_{\text{Edd}}$ did not show any correlation ($\uprho$ = -0.14, upper right panel of Fig. 6). A similar trend was observed in case of H$\alpha$ (lower panels of Fig. 6). Whereas almost no correlations were found in the relation between $\lambda_{\text{Edd}}$ and FWHM or EW of H$\beta$ and H$\alpha$ narrow component, implying that accretion onto the SMBH does not entirely affect the emission from the narrow line region.
\begin{figure}

	\includegraphics[width=\columnwidth]{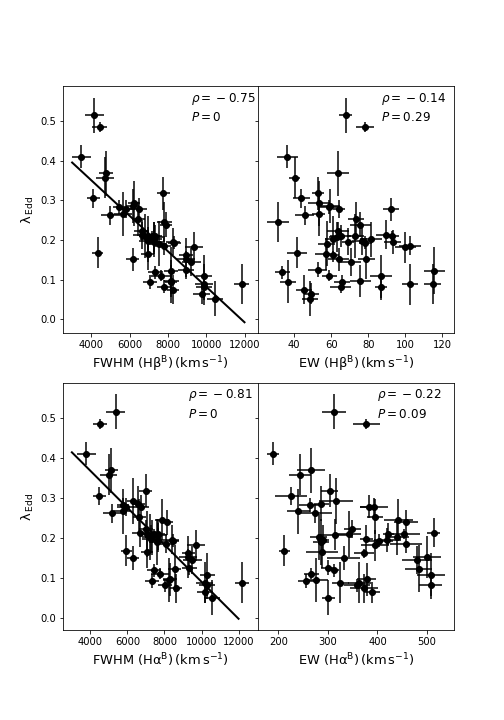}
    \caption{Eddington ratio vs FWHM (H${\beta}^{\text{B}}$) (left) and EW (H${\beta}^{\text{B}}$) (right). The solid line in each panel shows the linear best-fitting line.}
    \label{fig:6}
\end{figure}
\begin{figure*}
    \vspace*{-9mm}
	\includegraphics[width=\textwidth]{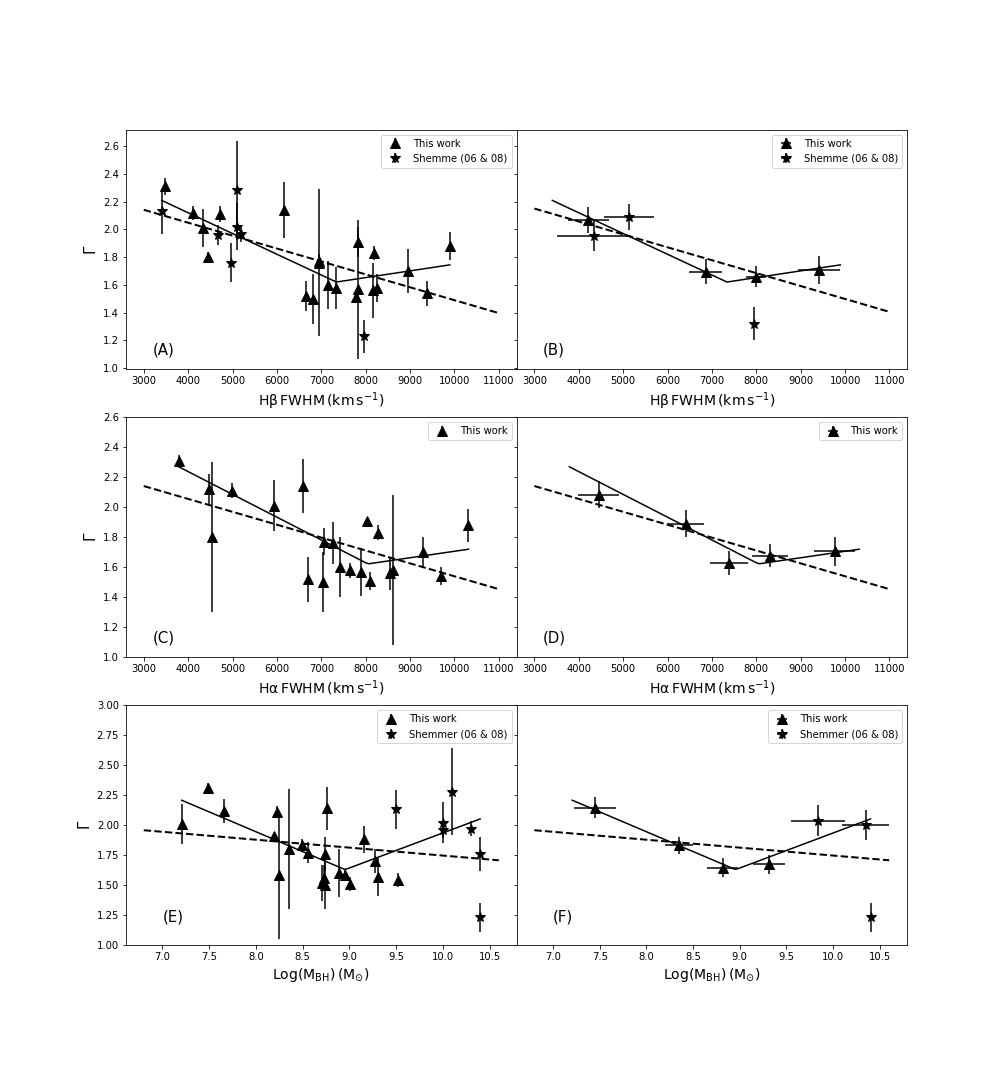}
    \caption{(A) X-ray photon index vs H$\beta$ FWHM. (C) X-ray photon index vs H$\alpha$ FWHM. (E): X-ray photon index vs M$_{\text{BH}}$.  (B), (D), and (F) represent the results for binned data to show clearly the locations of breaks.
    Data from different works are marked and shown in the panel. The solid and dashed lines represent the broken power-law best-fitting and linear best-fitting lines, respectively.}
    \label{fig:7}
\end{figure*}

\subsubsection {$\Gamma$ vs FWHM and M$_{\text{BH}}$}
The gravitational influence of SMBH that resides in the central region of AGN causes faster rotation of BLR in comparison with the rotation of NLR  (v$\approx(GM/R)^{1/2}$), where M is the mass of the SMBH and R is the distance of emitting region from the SMBH. Here we studied how FWHMs of broad components (ILR+VBLR) of H$\alpha$ and H$\beta$ and M$_{\text{BH}}$ are varying as a function of $\Gamma$.

The relations between $\Gamma$ and H$\beta$ FWHM have been reported by previous studies. Reeves \& Turner (2000) found an anti-correlation for a sample of 49 radio quiet AGNs and concluded that AGNs with the steepest X-ray spectra tend to have the narrowest H${\beta}$ FWHM. The same relation was observed by Laor et al. (1997) for a sample of 23 bright quasars,  Piconcelli et al. (2005) for a sample of 40 QSOs, and Porquet et al. (2004) for 21 type 1 AGNs. Recent study by Ojha et al. (2020) for a large sample of 375 (NLSy1+BLSy1) objects found that H${\beta}$ FWHM has an anti-correlation with X-ray photon index (0.1 - 2.0 keV), where they suggested that the high Eddington ratio associated with NLSy1 is responsible for the steeper $\Gamma$ compared to the BLSy1.\\

We obtained a negative correlation between $\Gamma$ and FWHM H${\beta}$ (Fig. 7A) which is similar to the results observed by Jin et al. (2012a) and Brightman et al. (2013). Interestingly, Jin et al. (2012a) observed a break around FWHM = 4000 km s$^{-1}$ in the study of $\Gamma$ vs FWHM H${\beta}$. In order to study further the break phenomenon, we included sources from Shemmer et al. (2006) and Shemmer et al. (2008) along with our data and found that the break location was shifted to $\sim$7336$\pm$650 km s$^{-1}$ (Fig. 7A). The astropy library (Astropy Collaboration et al. 2013, 2018) was used to compute the broken power law fits using Levenberg-Marquardt algorithm and least squares statistic for optimisation.
For the broken power-law fit $\chi^2$ = 0.66 and the relations are as follows.\\
For FWHM$_{\text{H}\beta}$  $<$ (7336$\pm$650) km s$^{-1}$  ($\uprho$ = -0.67, P = 0.002):
\begin{equation}
\Gamma=(-1.49\pm0.32)10^{-4} \text{FWHM}_{\text{H}\beta}+(2.71\pm0.18) 
\end{equation}
\\
For FWHM$_{\text{H}\beta}$  $>$ (7336$\pm$650) km s$^{-1}$ ($\uprho$ = 0.16, P = 0.66):
\begin{equation}
\Gamma=(4.82\pm2.40)10^{-4} \text{FWHM}_{\text{H}\beta}+(1.26\pm0.53)
\end{equation}
The linear best-fitting results for the whole sample are $\chi^2$= 0.86, $\uprho$ = -0.63, P = 0.0003 and the relation is as follow.
\begin{equation}
\Gamma=(-0.93\pm0.19) 10^{-4}\, \text{FWHM}_{\text{H}\beta}+(2.43\pm0.13) 
\end{equation}

The right panels of Fig. 7 show the result of binned data points in order to show the break position clearly. However, we argue that more data are need to confirm the break location especially at higher FWHM end.
We also noted a break in that relation between $\Gamma$ and FWHM of H$\alpha$ at FWHM of H$\alpha$=7642$\pm$657 km s$^{-1}$ (Fig. 7C). We fitted the broken power law ($\chi^2$= 0.56) as well as linear fit ($\chi^2$= 0.75) and found the following relations.\\
For FWHM$_{\text{H}\alpha}$  $<$ (7642$\pm$657) km s$^{-1}$  ($\uprho$ = -0.74, P = 0.008):
\begin{equation}
\Gamma=(-1.53\pm0.44)10^{-4} \text{FWHM}_{\text{H}\beta}+(2.79\pm0.27)
\end{equation}
\\
For FWHM$_{\text{H}\alpha}$  $>$ (7642$\pm$650) km s$^{-1}$ ($\uprho$ = 0.06, P = 0.85):
\begin{equation}
\Gamma=(0.43\pm1.6)10^{-4} \text{FWHM}_{\text{H}\beta}+(1.28\pm0.53)
\end{equation}
The linear best-fitting relation for the whole sample ($\uprho$ = -0.47, P = 0.027) is
\begin{equation}
\Gamma=(-0.86\pm0.25) 10^{-4}\, \text{FWHM}_{\text{H}\beta}+(2.40\pm0.19)
\end{equation}

Reverberation mapping method is the most accurate way to derive mass of SMBH in AGN, but because of the relatively small number of objects studied using this method, alternative ways of deriving masses are often used such as the relation between M$_{\text{BH}}$, H$\beta$ FWHM, and L$_{5100\text{\AA}}$ (Woo \& Urry 2002; Peterson et al. 2004; Jin et al. 2012c).

Using the relation given by Woo \& Urry (2002), we estimated masses of SMBHs
\begin{equation}
\text{M}_{\text{BH}} =  4.817 \Bigg[\frac{\lambda L_{\lambda}(5100\,\text{\AA})}{10^{44}\, \text{erg\,s}^{-1}}\Bigg]^{0.7}      \Big(\text{FWHM}_{\text{H}\beta}^{\text{B}}\Big)^{2}
\end{equation}

In the study of $\Gamma$ vs M$_{\text{BH}}$, a break was found at $\log$ (M$_{\text{BH}}$/M$_{\odot}$)=(8.95$\pm$0.21) (Fig. 7E; $\chi^2$ = 0.98 for a broken power-law fit and $\chi^2$ = 1.62 for a linear fit) giving the following relations.\\
For $\log$ (M$_{\text{BH}}$/M$_{\odot}$) $<$ (8.95$\pm$0.21) ($\uprho$ = -0.56, P = 0.023):
\begin{equation}
\Gamma=(-0.33\pm0.11) \log(\text{M}_{\text{BH}}/\text{M}_{\odot})+(4.59\pm0.90)
\end{equation}
For $\log$ (M$_{\text{BH}}$/M$_{\odot}$) $>$ (8.95$\pm$0.21) ($\uprho$ = 0.55, P = 0.06):
\begin{equation}
\Gamma=(0.29\pm0.13) \log(\text{M}_{\text{BH}}/\text{M}_{\odot})+(-0.99\pm1.22)
\end{equation}
The linear best-fitting relation for the whole sample ($\uprho$ = -0.08, P = 0.66) is 
\begin{equation}
\Gamma=(-0.07\pm0.06) \log(\text{M}_{\text{BH}}/\text{M}_{\odot})+(2.40\pm0.54)
\end{equation}

A similar study by Jin et al. (2012a) observed a break at $\log$ (M$_{\text{BH}}$/M$_{\odot}$)=8. 
Similar studies of the relation between $\Gamma$ and M$_{\text{BH}}$ in the future would be important to confirm the place of the break point.
It can be seen that as the M$_{\text{BH}}$ is increasing, the $\Gamma$ hardens until the break point but it softens again after the break. To explain this trend we suggest that in the place of the break, corona or Comptonizing region is relatively larger in size and after the break it is diminishing as it is experiencing more number of disc photons that lead to softening of $\Gamma$. The softening of $\Gamma$ could be due to two different scenarios, either the large values of FWHMs arising from the portion of BLR which is closer to the central SMBH, relatively increasing the soft photons flux and cooling the corona or it could be due to the increase in the inclination angle after the observed break, where corona is obscured by gas or dust causing the corona to appear softer. More studies are needed in this direction to confirm the above mentioned scenarios. As it can be seen that there is a variation in the break location, we suggest that more observations would be useful to accurately constrain the break location and the geometry of accretion disc. Since the Eddington ratio ($\lambda_{\text{Edd}}$) is correlated to the mass of SMBH as well as $\Gamma$ (Wang et al. 2004; Shemmer et al. 2006; Brightman et al. 2013), the $\lambda_{\text{Edd}}$ is also associated to the radio luminosity (e.g. Yang et al. 2020). This suggests that a break should be observed in the relation between $\lambda_{\text{Edd}}$ and radio emission parameter too.
Yang et al. (2020) noticed a break in the $\alpha$ (radio spectral index) vs $\lambda_{\text{Edd}}$ relation, associated with the sample of extremely high mass accretion rates.
It should be noted that Martocchia et al. (2017) reported a flatter power-law spectral index with a break at $\log$(M$_{\text{BH}}$/M$_{\odot}$)=(8.01$\pm$0.48), slope = -0.19 before the break and slope = 0.006 after the break, when compared to the slopes derived by Jin et al. (2012a), -0.37 before the break and 0.06 after the break.
The true picture can be understood if the number of data points is higher at lower and higher mass end. $\Gamma$ is closely related to the mass accretion rate due to thermal Compotnization in the corona and hence a weak or no correlation cannot be ruled out in $\Gamma$ vs M$_\text{BH}$ relation.\\
We observed a moderate anti-correlation between $\Gamma$ and EW of broad emission line components  for our sample (21 AGNs; Fig. 8). Harder $\Gamma$ was associated to the emission lines with a higher luminosity and we found no such correlation/anti-correlation with the narrow line components. 
\begin{figure}
	\includegraphics[width=\columnwidth]{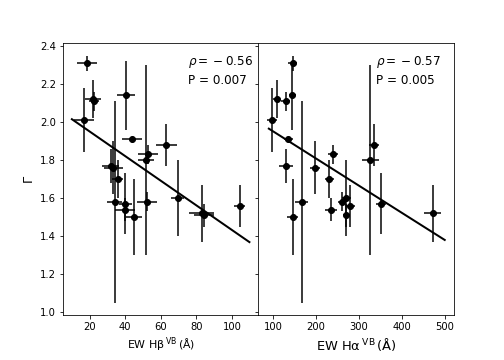}
    \caption{X-ray photon index vs EW (H${\beta}^{\text{VB}}$) (left) and EW (H${\alpha}^{\text{VB}}$) (right). The solid line in each panel shows the linear best-fitting line.}
    \label{fig:8}
\end{figure}
\subsection{Association of X-ray and MIR luminosities}
\begin{figure*}
	\includegraphics[width=\textwidth]{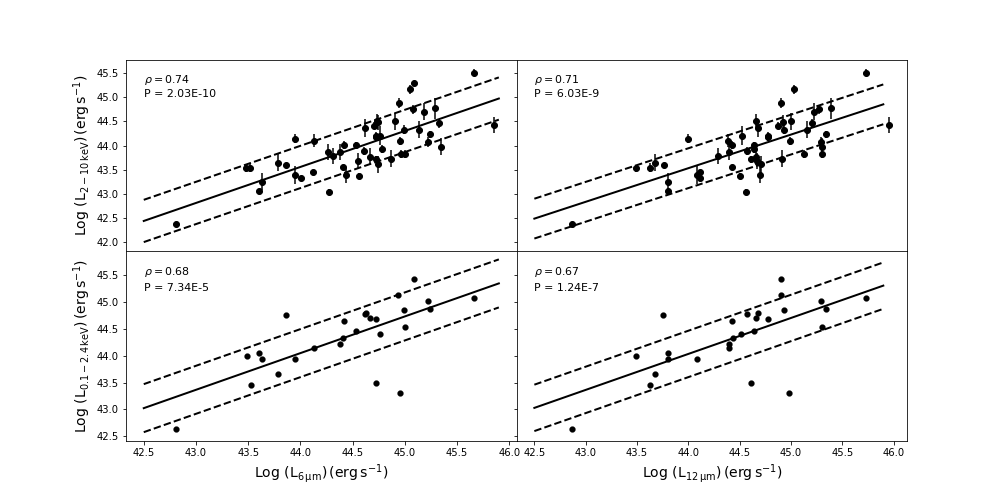}
    \caption{Upper panels: L$_{2-10\,\text{keV}}$ vs L$_{6\,\upmu\text{m}}$ and L$_{12\,\upmu \text{m}}$.
    Lower panels: L$_{0.1-2.4\,\text{keV}}$ vs L$_{6\,\upmu\text{m}}$ and L$_{12\,\upmu \text{m}}$. The solid line in each panel shows the regression line, the two dashed lines indicate the $\pm 1 \sigma$ region.}
    \label{fig:9}
\end{figure*}
Correlation between MIR and X-ray luminosities for AGNs is a well studied trend (e.g. Fiore et al. 2009; Gandhi et al. 2009; Asmus et al. 2015; Mateos et al. 2015; Garcia-Bernette et al. 2017). All these studies indicated a positive correlation, which is expected according to the unified emission model that the Compotonized X-ray emission in the corona is absorbed in torus and re-emitted in MIR as the continuum and emission lines. This relation is however subjected to many restrictions such as the obscuration by dust that changes depending on the structure and orientation of disc. Furthermore, the accretion disc luminosity also affects the relation because the radiation pressure produced by a higher disc luminosity pushes the dust away which in turn causes a lower obscuration. Our study, Fig. 9 (upper panels) show the correlations ($\uprho$ = 0.74, P = 2.03E-10 at L$_{6\,\upmu \text{m}}$ and $\uprho$ = 0.71, P = 6.03E-9 at L$_{12\,\upmu \text{m}}$ ) between MIR and X-ray (2-10 keV) luminosity for the selected sample.
The strength of correlation almost does not vary with wavelengths.
The correlation coefficient in this work is slightly lower than the one reported in Garcia-Bernette et al. (2017), $\uprho$ = 0.83 at L$_{6\,\upmu \text{m}}$ and $\uprho$ = 0.86 at L$_{12\,\upmu \text{m}}$ for a sample of 24 type 1 AGNs, whereas the coefficient is close to the value $\uprho$ = 0.7 at L$_{12\,\upmu \text{m}}$ obtained by Asmus et al. (2015) for a sample of 54 type 1 AGNs. Our best-fitting regression lines are as follows.
\begin{equation}
\log(L_{2-10\, \text{keV}})=(0.74\pm0.09) \log(L_{6\,\upmu \text{m}})+(10.81\pm4.12)
\end{equation}
\begin{equation}
\log(L_{2-10\, \text{keV}})=(0.70\pm0.09) \log(L_{12\,\upmu \text{m}})+(12.96\pm4.16)
\end{equation}

The slope of equation (24) is well in agreement with the study of Fiore et al. (2009) where the slope is about 0.72, but it is lower than the slope found by Garcia-Bernette et al. (2017) that is 0.99. Similarly we found the slope of equation (25) to be lower than the slopes reported by Gandhi et al. (2009) (0.90$\pm$0.06), Asmus et al. (2015) (1.04$\pm$0.05), and Garcia-Bernette et al. (2017) (0.86).
It is however necessary to note that from a large sample, Chen et al. (2017) found a bi-linear relation where the slopes change from 0.49 to 0.60 around the MIR luminosity 10$^{44.79}$ erg s$^{-1}$. Mateos et al. (2015) found a slope of about 0.81$\pm$0.06 for a sample of 103 type 1 AGNs with L$_{6\,\upmu \text{m}}$ $>$ 10$^{43.8}$ erg s$^{-1}$, similar to the obtained value in the present study.
The robustness of the slope is important because assuming a linear relationship L$_{\text{X}}$-L$_{6\,\upmu \text{m}}$ might result in an overestimation of column densities in the luminous AGNs (Chen et al. 2017).

A strong positive correlation was also observed between soft X-ray luminosity (0.1-2.4 keV) and MIR luminosity (Fig. 9 lower panels) and the regression lines slopes are shown below. 
\begin{equation}
 \log(L_{0.1-2.4\,\text{keV}})=(0.68\pm0.13) \log(L_{6\,\upmu \text{m}})+(14.02\pm6.11) 
\end{equation}
\begin{equation}
 \log(L_{0.1-2.4\,\text{keV}})=(0.67\pm0.14) \log(L_{12\,\upmu m})+(14.61\pm6.39) 
\end{equation}
The slopes are found to be almost same when compared to those of equation (24) and (25). This result indicates that the MIR emissions at the torus are similarly affected by both soft and hard X-rays.
We studied the correlation between $\Gamma$ and MIR luminosities at 6 $\upmu$m and 12 $\upmu$m.
We found that $\Gamma$ is negatively correlated with L$_{6\,\upmu \text{m}}$ ($\uprho$ =  -0.49, P = 0.021), but the correlation is not valid in case of L$_{12\,\upmu \text{m}}$ where P = 0.058 > 0.05 as shown in Fig. 10 and equations in (28) and (29).
\begin{equation}
 \Gamma=(-0.21\pm0.06) \log(L_{6\,\upmu \text{m}})+(11.38\pm2.98) 
\end{equation}
\begin{equation}
 \Gamma=(-0.19\pm0.06) \log(L_{12\,\upmu \text{m}})+(10.45\pm3.08)
\end{equation}
\begin{figure}
	\includegraphics[width=\linewidth]{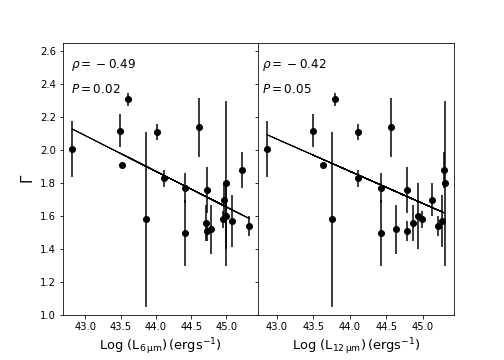}
    \caption{X-ray photon index vs L$_{6\,\upmu \text{m}}$ and L$_{12\,\upmu \text{m}}$. The solid line in each panel shows the linear best-fitting line.}
    \label{fig:10}
\end{figure}

 It clearly suggests that the MIR luminosities are affected by the X-rays i.e. higher MIR luminosities are associated with hard $\Gamma$ and similar correlations are also observed between $\Gamma$ and H$\alpha$, H$\beta$ emissions. This outcome of the correlations could be due to the fact that the harder $\Gamma$ is often associated to the physically larger corona above the accretion disc which in turn provides more X-ray photons to the torus and in effect producing the higher MIR emission. We also noted that spectral index of MIR is not correlated with any of the parameters viz. $\Gamma$, $\lambda_{\text{Edd}}$, L$_{\text{X}}$, FWHM, and EW of H${\alpha}$ and H${\beta}$ components Lakicevic et al. (2017). 

\subsection{Balmer decrements}
Balmer decrements have been used in many studies as an indicator of reddening in different regions of AGNs  (e.g. La Mura et al. 2007; Dong et al. 2008; Jin et al. 2012b; Gaskell 2017). Previous studies showed that narrow and broad regions have different Balmer decrements, indicating that fluxes of emission lines originating from these regions are subjected to different degree of reddening. In addition, the value of the decrement slightly changes from one sample to another, for example, Lu et al. (2019) case is H$\alpha^{\text{B}}$/H$\beta^{\text{B}}$ = 3.16, H$\alpha^{\text{N}}$/H$\beta^{\text{N}}$ = 4.37, Gaskell (2017) H$\alpha^{\text{B}}$/H$\beta^{\text{B}}$ = 2.72, Gaskell (1982) H$\alpha^{\text{N}}$/H$\beta^{\text{N}}$ = 3.1, and Dong et al. (2008) H$\alpha^{\text{B}}$/H$\beta^{\text{B}}$ = 3.06 with a standard deviation 0.03 dex. We calculated Balmer decrements for NLR, ILR, VBLR, and BLR and obtained the mean values to be 3.75 $\pm$ 0.01, 4.46 $\pm$ 0.01, 2.91 $\pm$ 0.05, and 3.38 $\pm$ 0.02 for NLR, ILR, VBLR, BLR, respectively. These are similar to the values obtained by Lakicevic et al. (2017). \\
Here we studied the relation between broad Balmer decrement and $\Gamma$, $\lambda_{\text{Edd}}$. The result, in Fig. 11 shows that the Balmer decrement is increasing along with the hardening of $\Gamma$.
 $\Gamma$ is associated with the corona/jet and the inner region of the accretion disc. The corona has been found to be compact with a high optical depth in case of steep $\Gamma$ (e.g. Haardt \& Maraschi 1991; Haardt \& Maraschi 1993; Czerny et al. 2003) and the configuration of disc-corona geometry changes with a mass accretion rate (e.g. Kelly et al. 2008). From our sample, it has been found that the Balmer decrement increases along with the decrease of Eddington ratio (Fig. 12). The connection between Balmer decrement and $\Gamma$, $\lambda_{\text{Edd}}$ can be explained in the following scenario, i.e. some of the material associated with the dust from broad line region settles on the outer region of the accretion disc and traverses closer to the inner region, affecting the $\Gamma$. 
The steeper $\Gamma$ of corona is also associated with a higher accretion rate where the Balmer decrement is low (Fig. 12).

\begin{figure}
	\includegraphics[width=\columnwidth]{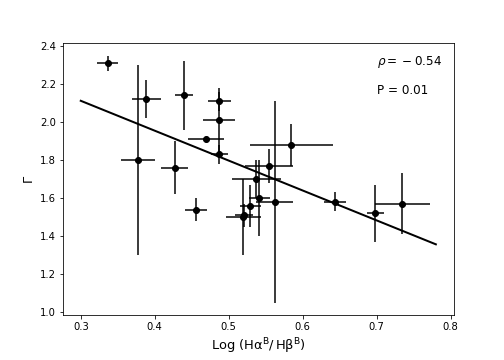}
    \caption{X-ray photon index vs broad Balmer decrement. The solid line shows the linear best-fitting line.}
    \label{fig:11}
\end{figure}

This fact can be explained if most of the dust/gas or any material which is causing the decrement is located along the accretion disc plane, which causes less obscuration. Whenever the Balmer decrement is high, the accretion rate is low and $\Gamma$ is relatively harder. In this scenario, most of the decrement causing material is located above the disc plane. This scenario can also explain the evolution of Balmer decrement in a sense that the dust/gas or obscuring material above the disc's plane settles down over the disc which causes the increase in the mass accretion. As the accreting material approaches closer to the inner region of the accretion disc, the disc in turn emits larger number of  soft photons (optical/UV) that  causes cooling of the corona. Such a physical scenario was invoked to explain the evolution of covering factor of circumnuclear material vs the mass accretion rate (Ricci et al. 2017). The duration taken by the obscuring material above the disc's plane to settle down along the primary accretion flow is not known, however, it has been suggested by Ricci et al. (2017) that short timescales would be required to clear out the material. Contrary to our finding, Dong et al. (2008) did not find any correlation between broad Balmer decrement and Eddington ratio for a sample of 446 low redshift blue AGNs. On the other hand, La Mura et al. (2007) studied the relation for a sample of 90 broad line AGNs and reported a weak anti-correlation that was explained as a result of the high reddening for low accretion rate objects in their sample. Similarly, Lu et al. (2019) suggested this anti-correlation for a sample of 554 AGNs to be an evidence that the accretion rate affects the reddening of BLR.
\begin{figure}
	\includegraphics[width=\columnwidth]{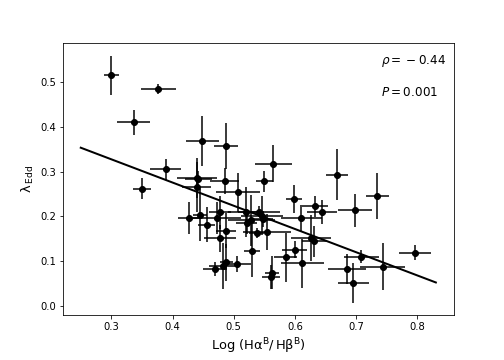}
    \caption{Eddington ratio vs broad Balmer decrement. The solid line shows the linear best-fitting line.}
    \label{fig:12}
\end{figure}\\
\section{Conclusions}
We extracted a sample of type 1 AGNs from the SDSS data base DR12 and reduced and analysed the SDSS data independently by adhering all the methods and procedures. Our focus was to evaluate the relations X-ray vs optical and X-ray vs MIR band which cover a broad span of various parameters and understand their physical meaning. Our new findings and analysis resulted in the following conclusions.

\begin{enumerate}

 \item Different components of Balmer emission lines were found to have different degree of correlation with X-ray luminosities (2-10 keV), where the tightest correlations exist in the VBLR components while the lowest correlations were associated to the NLR components. This fact strongly suggests that the VBLR component is closer to the central engine. The slopes ($\sim$ 0.74) of regression lines derived in this study were very close to the slopes obtained by Jin et al. (2012b). It was also confirmed that our other correlations/slopes such as L$_{\text{X}}$ vs L$_{5100\,\text{\AA}}$, L$_{\text{X}}$ vs L$_{\text{MIR}}$, were in good agreement with those of previous other studies.  

 \item We found that the broader components of Balmer lines tend to have a lower Eddington ratio, while the brighter lines (i.e. EWs) did not show such similar trend. This result indicates that the mass accretion rate is key parameter in regulating and modulating the BLR region. Such a correlation was not observed for the NLR region.

 \item Breaks were found in the relations between $\Gamma$ -- FWHM H${{\beta}^\text{B}}$ at FWHM$_{\text{H}\beta}$  = (7336$\pm$650) km s$^{-1}$ and $\Gamma$ -- FWHM H${{\alpha}^\text{B}}$ at FWHM$_{\text{H}\alpha}$  = (7642$\pm$657) km s$^{-1}$. Similarly in the relation between $\Gamma$ and M$_{\text{BH}}$ a break was found at $\log$ (M$_{\text{BH}}$/M$_{\odot}$)$=$(8.95$\pm$0.21). Such a break phenomenon is supported by the radio observations (Yang et al. 2020). However, more data would be helpful to properly constrain the breaks locations and its physical significance.

 \item A tight correlation exists between MIR and X-ray luminosity (2-10 keV). The values of obtained slopes are consistent with the results of Mateos et al. (2015) and Fiore et al. (2009) but, slightly differ from the studies of Gandhi et al. (2009), Asmus et al. (2015), and Garcia-Bernette et al.  (2017). It is noteworthy that there is also a tight correlation between soft X-ray (0.1-2.4 keV) and MIR luminosities.

 \item $\Gamma$ was negatively correlated with the MIR luminosity. This result suggests that a physically larger corona may be responsible for the higher MIR emission from the torus sturcture in AGNs.

\item Balmer decrements were found to be increasing along with hardening of the $\Gamma$ as well as such similar trend was noticed with respect to $\lambda_{\text{Edd}}$. This indicates that AGN with a physically large corona hosts relatively more obscuring material in the BLR region.
\end{enumerate}

\begin{table*}
\caption{H$\beta$ spectral parameters derived from the fit.}
\label{tab:T1}
\setlength\tabcolsep{3 pt}
\setlength{\cmidrulekern}{2em}
\scriptsize
\centering
\begin{tabular}{|l|c|c|c|c|ccc|ccc|ccc|}

\hline

 &  &  &  &  & \multicolumn{3}{c|}{H$\beta$ EW ($\text{\AA}$)} & \multicolumn{3}{c|}{H$\beta$ FWHM (km s$^{-1}$)} &\multicolumn{3}{c|}{Log L$_{\text{H}\beta}$ (erg s$^{-1}$) }\\
 \cline{6-8} \cline{9-11} \cline{12-14}
ID & SDSS name & RA & Dec & z &
NLR & ILR & VBLR & NLR & ILR & VBLR & NLR & ILR & VBLR \\
 (1) & (2) & (3) & (4) & (5) & (6) & (7) & (8) & (9) & (10) & (11) & (12) & (13) & (14)\\

\hline
1 & J005812.85+160201.37 & 00 58 12.84 & +16 02 01.41 & 0.21 & 0.98$\pm$0.38 & 21.03$\pm$1.07 & 43.28$\pm$3.39 & 364$\pm$29 & 1852$\pm$124 & 5505$\pm$462 & 40.99$\pm$0.09 & 42.32$\pm$0.08 & 42.63$\pm$0.05 \\
2 & J015950.24+002340.91 & 01 59 50.25 & +00 23 40.87 & 0.16 & 3.92$\pm$0.28 & 26.68$\pm$2.91 & 51.66$\pm$4.73 & 501$\pm$32 & 1913$\pm$56 & 4010$\pm$381 & 41.50$\pm$0.14 & 42.34$\pm$0.07 & 42.62$\pm$0.07 \\
3 & J020615.99-001729.20 & 02 06 15.98 & -00 17 29.21 & 0.04 & 1.12$\pm$0.51 & 11.26$\pm$1.61 & 34.15$\pm$4.21 & 397$\pm$33 & 3178$\pm$165 & 7627$\pm$312 & 40.01$\pm$0.14 & 41.01$\pm$0.04 & 41.48$\pm$0.09 \\
4 & J021447.00-003250.62 & 02 14 47.00 & -00 32 50.64 & 0.35 & 3.35$\pm$0.59 & 15.28$\pm$1.84 & 21.54$\pm$4.32 & 423$\pm$25 & 4099$\pm$145 & 7058$\pm$324 & 40.91$\pm$0.08 & 41.57$\pm$0.06 & 41.72$\pm$0.06 \\
5 & J073623.13+392617.81 & 07 36 23.12 & +39 26 17.70 & 0.12 & 5.46$\pm$0.65 & 67.03$\pm$1.37 & 47.71$\pm$4.45 & 509$\pm$44 & 2645$\pm$77 & 11559$\pm$398 & 41.41$\pm$0.11 & 42.49$\pm$0.11 & 42.34$\pm$0.14 \\
6 & J081652.25+425829.42 & 08 16 52.24 & +42 58 29.41 & 0.23 & 0.89$\pm$0.12 & 27.75$\pm$2.99 & 45.64$\pm$4.18 & 295$\pm$28 & 2097$\pm$181 & 6113$\pm$348 & 40.88$\pm$0.05 & 42.38$\pm$0.07 & 42.59$\pm$0.07 \\
7 & J091955.35+552137.12 & 09 19 55.34 & +55 21 37.12 & 0.12 & 1.95$\pm$0.55 & 18.63$\pm$2.82 & 62.99$\pm$5.87 & 384$\pm$33 & 1875$\pm$195 & 7044$\pm$397 & 41.14$\pm$0.09 & 42.12$\pm$0.04 & 42.65$\pm$0.11 \\
8 & J094636.43+205610.98 & 09 46 36.42 & +20 56 10.89 & 0.28 & 1.54$\pm$0.18 & 38.44$\pm$2.75 & 40.06$\pm$3.50 & 537$\pm$41 & 3699$\pm$177 & 7462$\pm$349 & 41.08$\pm$0.14 & 42.48$\pm$0.05 & 42.49$\pm$0.04 \\
9 & J095048.39+392650.52 & 09 50 48.38 & +39 26 50.46 & 0.21 & 1.37$\pm$0.14 & 23.19$\pm$1.03 & 69.42$\pm$3.81 & 403$\pm$22 & 2251$\pm$135 & 6785$\pm$415 & 41.19$\pm$0.08 & 42.42$\pm$0.12 & 42.89$\pm$0.10 \\
10 & J100420.14+051300.43 & 10 04 20.13 & +05 13 00.39 & 0.16 & 7.23$\pm$0.62 & 29.70$\pm$1.89 & 38.07$\pm$3.46 & 1071$\pm$25 & 1878$\pm$218 & 3695$\pm$496 & 41.76$\pm$0.15 & 42.37$\pm$0.11 & 42.48$\pm$0.04 \\
11 & J100642.59+412201.88 & 10 06 42.581 & +41 22 01.95  & 0.15 & 6.27$\pm$0.28 & 12.58$\pm$2.79 & 35.85$\pm$4.35 & 608$\pm$42 & 2675$\pm$58 & 10088$\pm$414 & 41.36$\pm$0.12 & 41.66$\pm$0.15 & 42.11$\pm$0.12 \\
12 & J100726.10+124856.22  & 10 07 26.09 & +12 48 56.18 & 0.24 & 1.94$\pm$0.60 & 22.36$\pm$2.96 & 9.08$\pm$5.97 & 327$\pm$57 & 4065$\pm$174 & 6676$\pm$377 & 41.83$\pm$0.09 & 42.89$\pm$0.08 & 42.47$\pm$0.05 \\
13 & J102531.28+514034.87 & 10 25 31.27 & +51 40 34.86 & 0.04 & 0.84$\pm$0.54 & 22.19$\pm$2.86 & 21.78$\pm$4.68 & 357$\pm$31 & 1426$\pm$86 & 3854$\pm$344 & 39.90$\pm$0.12 & 41.32$\pm$0.06 & 41.32$\pm$0.05 \\
14 & J105143.90+335926.70 & 10 51 43.89 & +33 59 26.68 & 0.17 & 1.12$\pm$0.68 & 28.07$\pm$2.89 & 44.93$\pm$4.47 & 329$\pm$25 & 2613$\pm$95 & 6285$\pm$312 & 40.95$\pm$0.05 & 42.35$\pm$0.14 & 42.55$\pm$0.13 \\
15 & J105151.45-005117.68 & 10 51 51.44 &  -00 51 17.70 & 0.36 & 2.23$\pm$0.33 & 6.71$\pm$1.31 & 62.33$\pm$3.41 & 378$\pm$22 & 1446$\pm$194 & 8152$\pm$361 & 42.28$\pm$0.07 & 42.76$\pm$0.13 & 43.72$\pm$0.05 \\
16 & J105705.41+580437.52 & 10 57 05.40 & +58 04 37.45 & 0.14 & 2.07$\pm$0.41 & 36.39$\pm$2.18 & 29.47$\pm$4.24 & 233$\pm$30 & 2890$\pm$186 & 6443$\pm$366 & 40.32$\pm$0.11 & 41.57$\pm$0.11 & 41.47$\pm$0.14 \\
17 & J111706.40+441333.31 & 11 17 06.39 & +44 13 33.32 & 0.14 & 1.22$\pm$0.30 & 12.86$\pm$2.68 & 52.36$\pm$5.71 & 410$\pm$23 & 2295$\pm$198 & 6957$\pm$389 & 41.20$\pm$0.04 & 42.22$\pm$0.07 & 42.83$\pm$0.07 \\
18 & J111830.29+402554.01 & 11 18 30.28 & +40 25 54.06 & 0.15 & 0.92$\pm$0.22 & 19.18$\pm$1.00 & 40.30$\pm$5.06 & 427$\pm$46 & 1348$\pm$194 & 6010$\pm$348 & 41.03$\pm$0.06 & 42.35$\pm$0.06 & 42.67$\pm$0.07 \\
19 & J111908.67+211917.9 & 11 19 08.67 & +21 19 17.98 & 0.18 & 2.97$\pm$0.47 & 60.15$\pm$2.77 & 39.73$\pm$5.71 & 827$\pm$39 & 2705$\pm$201 & 8972$\pm$472 & 42.12$\pm$0.12 & 43.42$\pm$0.07 & 43.24$\pm$0.15 \\
20 & J112439.18+420145.03 & 11 24 39.17 & +42 01 45.01 & 0.22 & 2.05$\pm$0.42 & 48.05$\pm$2.27 & 44.37$\pm$4.27 & 479$\pm$20 & 2341$\pm$55 & 6030$\pm$443 & 41.62$\pm$0.11 & 42.99$\pm$0.11 & 42.96$\pm$0.14 \\
21 & J114043.62+532439.03 & 11 40 43.62 & +53 24 38.96 & 0.53 & 1.69$\pm$0.56 & 24.77$\pm$2.90 & 27.96$\pm$5.06 & 396$\pm$23 & 4035$\pm$93 & 7999$\pm$409 & 41.31$\pm$0.08 & 42.48$\pm$0.07 & 42.53$\pm$0.13 \\
22 & J115324.47+493108.76 & 11 53 24.46 & +49 31 08.83 & 0.33 & 1.48$\pm$0.16 & 26.93$\pm$2.36 & 36.56$\pm$5.98 & 397$\pm$24 & 2905$\pm$166 & 5954$\pm$378 & 41.18$\pm$0.09 & 42.43$\pm$0.08 & 42.57$\pm$0.10 \\
23 & J115341.78+461242.25 & 11 53 41.77 & +46 12 42.17 & 0.02 & 4.74$\pm$0.36 & 24.89$\pm$2.66 & 16.71$\pm$5.56 & 256$\pm$45 & 873$\pm$182 & 4236$\pm$306 & 39.93$\pm$0.10 & 40.65$\pm$0.12 & 40.48$\pm$0.15 \\
24 & J120442.11+275411.82 & 12 04 42.10 & +27 54 11.86 & 0.17 & 4.51$\pm$0.23 & 11.26$\pm$2.77 & 88.30$\pm$5.68 & 398$\pm$22 & 1189$\pm$85 & 8075$\pm$370 & 41.34$\pm$0.13 & 41.74$\pm$0.08 & 42.70$\pm$0.04 \\
25 & J122539.56+245836.39 & 12 25 39.54 & +24 58 36.38 & 0.27 & 3.59$\pm$0.43 & 23.65$\pm$1.61 & 63.18$\pm$3.01 & 395$\pm$25 & 3930$\pm$156 & 9079$\pm$425 & 41.59$\pm$0.10 & 42.41$\pm$0.12 & 42.84$\pm$0.13 \\
26 & J123203.63+200929.55 & 12 32 03.62 & +20 09 29.49 & 0.06 & 1.24$\pm$0.42 & 22.79$\pm$2.99 & 52.97$\pm$5.71 & 309$\pm$23 & 2289$\pm$164 & 7858$\pm$378 & 40.40$\pm$0.06 & 41.67$\pm$0.07 & 42.03$\pm$0.09 \\
27 & J124635.25+022208.75 & 12 46 35.25 & +02 22 08.78 & 0.05 & 4.00$\pm$0.58 & 17.72$\pm$2.11 & 18.57$\pm$5.54 & 403$\pm$41 & 689$\pm$136 & 3416$\pm$498 & 39.54$\pm$0.07 & 41.19$\pm$0.11 & 41.21$\pm$0.15 \\
28 & J130947.00+081948.24 & 13 09 47.00 & +08 19 48.21 & 0.15 & 2.29$\pm$0.33 & 18.18$\pm$2.64 & 84.32$\pm$5.68 & 372$\pm$47 & 1779$\pm$52 & 7564$\pm$369 & 41.50$\pm$0.08 & 42.40$\pm$0.08 & 43.06$\pm$0.07 \\
29 & J131217.75+351521.06 & 13 12 17.75 & +35 15 21.08 & 0.18 & 1.70$\pm$0.17 & 25.01$\pm$2.63 & 35.81$\pm$3.17 & 529$\pm$34 & 3767$\pm$78 & 8116$\pm$368 & 41.54$\pm$0.06 & 42.71$\pm$0.07 & 42.86$\pm$0.15 \\
30 & J134356.75+253847.69 & 13 43 56.74 & +25 38 47.63 & 0.09 & 0.94$\pm$0.40 & 33.25$\pm$2.83 & 30.70$\pm$5.51 & 407$\pm$38 & 2806$\pm$208 & 5487$\pm$337 & 40.26$\pm$0.12 & 41.81$\pm$0.05 & 41.77$\pm$0.12 \\
31 & J135550.21+204614.54 & 13 55 50.20 & +20 46 14.51 & 0.20 & 8.76$\pm$0.19 & 28.36$\pm$1.20 & 25.06$\pm$5.84 & 1097$\pm$57 & 2451$\pm$177 & 5726$\pm$333 & 42.21$\pm$0.14 & 42.72$\pm$0.09 & 42.67$\pm$0.11 \\
32 & J135632.80+210352.40 & 13 56 32.79 & +21 03 52.36 & 0.30 & 1.54$\pm$0.64 & 20.90$\pm$1.52 & 58.03$\pm$4.51 & 398$\pm$24 & 3251$\pm$105 & 8318$\pm$491 & 41.39$\pm$0.08 & 42.53$\pm$0.09 & 42.97$\pm$0.10 \\
33 & J140251.20+263117.59 & 14 02 51.19 & +26 31 17.55 & 0.19 & 0.62$\pm$0.45 & 6.61$\pm$2.17 & 82.88$\pm$5.07 & 502$\pm$30 & 1129$\pm$177 & 6570$\pm$466 & 40.64$\pm$0.13 & 41.66$\pm$0.08 & 42.76$\pm$0.04 \\
34 & J140438.80+432707.43 & 14 04 38.79 & +43 27 07.45 & 0.32 & 0.59$\pm$0.27 & 26.72$\pm$1.58 & 26.11$\pm$3.21 & 845$\pm$59 & 2344$\pm$82 & 7407$\pm$345 & 41.65$\pm$0.11 & 43.31$\pm$0.11 & 43.29$\pm$0.06 \\
35 & J141556.85+052029.56 & 14 15 56.83 & +05 20 29.58 & 0.13 & 1.53$\pm$0.12 & 24.94$\pm$2.33 & 24.38$\pm$4.23 & 267$\pm$29 & 3202$\pm$106 & 9219$\pm$426 & 40.49$\pm$0.10 & 41.70$\pm$0.11 & 41.69$\pm$0.07 \\
36 & J141700.82+445606.36 & 14 17 00.82 & +44 56 06.33 & 0.11 & 0.77$\pm$0.34 & 25.33$\pm$1.36 & 31.74$\pm$5.68 & 452$\pm$44 & 2239$\pm$169 & 6750$\pm$324 & 40.48$\pm$0.08 & 41.99$\pm$0.06 & 42.09$\pm$0.09 \\
37 & J142748.29+050222.04 & 14 27 48.28 & +05 02 22.04 & 0.11 & 4.38$\pm$0.16 & 28.74$\pm$2.51 & 30.19$\pm$5.25 & 478$\pm$54 & 1313$\pm$118 & 5297$\pm$321 & 41.32$\pm$0.07 & 42.14$\pm$0.06 & 42.16$\pm$0.07 \\
38 & J142943.07+474726.20 & 14 29 43.07 & +47 47 26.21 & 0.22 & 2.73$\pm$0.54 & 43.26$\pm$2.48 & 33.34$\pm$5.39 & 441$\pm$33 & 2333$\pm$179 & 6555$\pm$474 & 41.62$\pm$0.09 & 42.82$\pm$0.06 & 42.71$\pm$0.13 \\
39 & J144645.94+403505.76 & 14 46 45.93 & +40 35 05.79 & 0.27 & 3.59$\pm$0.65 & 23.66$\pm$1.51 & 63.17$\pm$5.95 & 395$\pm$39 & 3931$\pm$112 & 9080$\pm$424 & 41.59$\pm$0.05 & 42.41$\pm$0.08 & 42.84$\pm$0.08 \\
40 & J144825.10+355946.65 & 14 48 25.09 & +35 59 46.70 & 0.11 & 1.61$\pm$0.29 & 30.03$\pm$2.56 & 29.04$\pm$4.13 & 366$\pm$49 & 1768$\pm$57 & 7408$\pm$487 & 40.66$\pm$0.13 & 41.93$\pm$0.05 & 41.91$\pm$0.14 \\
41 & J145108.76+270926.92 & 14 51 08.76 & +27 09 26.96 & 0.06 & 2.51$\pm$0.69 & 17.75$\pm$1.58 & 22.52$\pm$3.03 & 269$\pm$37 & 893$\pm$83 & 4639$\pm$457 & 41.02$\pm$0.13 & 41.86$\pm$0.04 & 41.97$\pm$0.06 \\
42 & J145608.65+275008.75 & 14 56 08.65 & +27 50 08.71 & 0.25 & 0.78$\pm$0.31 & 32.74$\pm$1.81 & 30.48$\pm$4.23 & 564$\pm$37 & 2130$\pm$98 & 6940$\pm$497 & 40.96$\pm$0.07 & 42.58$\pm$0.06 & 42.55$\pm$0.07 \\
43 & J152114.26+222743.87 & 15 21 14.25 & +22 27 43.82 & 0.14 & 11.65$\pm$0.32 & 57.93$\pm$1.29 & 44.59$\pm$4.25 & 1178$\pm$48 & 2591$\pm$66 & 9519$\pm$466 & 41.84$\pm$0.04 & 42.54$\pm$0.10 & 42.42$\pm$0.06 \\
44 & J153552.40+575409.50 & 15 35 52.40 & +57 54 09.51 & 0.03 & 3.11$\pm$0.61 & 21.28$\pm$2.76 & 43.84$\pm$5.55 & 387$\pm$57 & 2903$\pm$191 & 7257$\pm$371 & 40.45$\pm$0.04 & 41.28$\pm$0.11 & 41.59$\pm$0.04 \\
45 & J154307.78+193751.75 & 15 43 07.77 & +19 37 51.74 & 0.23 & 1.94$\pm$0.17 & 24.54$\pm$2.14 & 68.62$\pm$4.75 & 438$\pm$41 & 3025$\pm$203 & 6598$\pm$480 & 41.20$\pm$0.10 & 42.30$\pm$0.07 & 42.75$\pm$0.11 \\
46 & J154743.54+205216.66 & 15 47 43.53 & +20 52 16.61 & 0.26 & 2.38$\pm$0.35 & 9.93$\pm$2.69 & 65.83$\pm$5.58 & 427$\pm$42 & 2244$\pm$197 & 7580$\pm$306 & 41.92$\pm$0.05 & 42.54$\pm$0.14 & 43.36$\pm$0.10 \\
47 & J155444.58+082221.45 & 15 54 44.57 & +08 22 21.41 & 0.12 & 4.49$\pm$0.35 & 19.84$\pm$2.15 & 33.51$\pm$3.45 & 538$\pm$37 & 1666$\pm$147 & 5400$\pm$460 & 41.50$\pm$0.14 & 42.15$\pm$0.08 & 42.37$\pm$0.10 \\
48 & J155936.14+544203.77 & 15 59 36.14 & +54 42 03.80 & 0.31 & 1.42$\pm$0.28 & 25.35$\pm$1.84 & 38.20$\pm$5.87 & 266$\pm$55 & 1647$\pm$91 & 4454$\pm$362 & 40.88$\pm$0.10 & 42.13$\pm$0.12 & 42.31$\pm$0.11 \\
49 & J161413.20+260416.20 & 16 14 13.20 & +26 04 16.22 & 0.13 & 4.58$\pm$0.38 & 19.95$\pm$2.49 & 37.64$\pm$4.44 & 352$\pm$32 & 1959$\pm$115 & 7257$\pm$437 & 41.75$\pm$0.07 & 42.39$\pm$0.07 & 42.66$\pm$0.05 \\
50 & J163459.83+204936.09 & 16 34 59.82 & +20 49 36.12 & 0.13 & 1.67$\pm$0.42 & 11.56$\pm$1.57 & 22.15$\pm$4.09 & 823$\pm$45 & 3062$\pm$165 & 6680$\pm$392 & 40.49$\pm$0.08 & 41.33$\pm$0.09 & 41.62$\pm$0.08 \\
51 & J171207.44+584754.51 & 17 12 07.43 & +58 47 54.45 & 0.27 & 4.05$\pm$0.13 & 26.05$\pm$1.96 & 35.11$\pm$4.67 & 589$\pm$38 & 1788$\pm$71 & 6726$\pm$434 & 41.50$\pm$0.11 & 42.31$\pm$0.11 & 42.44$\pm$0.12 \\
52 & J223607.68+134355.3 & 22 36 07.68 & +13 43 55.36 & 0.33 & 0.87$\pm$0.49 & 16.55$\pm$2.86 & 29.09$\pm$4.94 & 304$\pm$58 & 1355$\pm$83 & 4813$\pm$483 & 41.57$\pm$0.09 & 42.84$\pm$0.13 & 43.09$\pm$0.06 \\
53 & J235156.13-010913.35 & 23 51 56.12 & -01 09 13.31 & 0.17 & 1.83$\pm$0.61 & 23.02$\pm$1.63 & 47.61$\pm$5.59 & 408$\pm$47 & 3573$\pm$131 & 8498$\pm$497 & 41.36$\pm$0.05 & 42.46$\pm$0.06 & 42.77$\pm$0.14 \\

\hline
\end{tabular}
\end{table*}

\begin{table*}
\caption{H$\alpha$ spectral parameters derived from the fit.}
\label{tab:T2}
\setlength\tabcolsep{3 pt}
\setlength{\cmidrulekern}{2 em}
\scriptsize
\begin{tabular}{|l|c|c|c|c|ccc|ccc|ccc|}

\hline

 &  &  &  &  & \multicolumn{3}{c|}{H$\alpha$ EW ($\text{\AA}$)} & \multicolumn{3}{c|}{H$\alpha$ FWHM (km s$^{-1}$)} &\multicolumn{3}{c|}{Log L$_{\text{H}\alpha}$ (erg s$^{-1}$) }\\
 \cline{6-9} \cline{9-11} \cline{12-14}
ID & SDSS name & RA & Dec & z &
NLR & ILR & VBLR & NLR & ILR & VBLR & NLR & ILR & VBLR \\
 (1) & (2) & (3) & (4) & (5) & (6) & (7) & (8) & (9) & (10) & (11) & (12) & (13) & (14)\\
\hline
1 & J005812.85+160201.37 & 00 58 12.84 & +16 02 01.41 & 0.21 & 7.07$\pm$0.74 & 163.20$\pm$3.32 & 229.21$\pm$8.04 & 362$\pm$29 & 2426$\pm$36 & 7804$\pm$137 & 41.74$\pm$0.07 & 43.10$\pm$0.17 & 43.25$\pm$0.17 \\
2 & J015950.24+002340.91 & 01 59 50.25 & +00 23 40.87 & 0.16 & 6.61$\pm$0.89 & 51.54$\pm$2.46 & 326.26$\pm$5.49 & 499$\pm$11 & 1399$\pm$54 & 6315$\pm$66 & 41.43$\pm$0.02 & 42.33$\pm$0.16 & 43.12$\pm$0.16 \\
3 & J020615.99-001729.20 & 02 06 15.98 & -00 17 29.21 & 0.04 & 11.30$\pm$0.54 & 205.78$\pm$6.39 & 167.01$\pm$14.56 & 395$\pm$9 & 3460$\pm$64 & 8609$\pm$321 & 40.66$\pm$0.16 & 41.91$\pm$0.18 & 41.82$\pm$0.21 \\
4 & J021447.00-003250.62 & 02 14 47.00 & -00 32 50.64 & 0.35 & 45.81$\pm$5.88 & 165.09$\pm$22.09 & 110.64$\pm$70.16 & 421$\pm$47 & 3289$\pm$416 & 7250$\pm$1645 & 41.78$\pm$0.04 & 42.34$\pm$0.07 & 42.17$\pm$0.06 \\
5 & J073623.13+392617.81 & 07 36 23.12 & +39 26 17.70 & 0.12 & 14.89$\pm$1.05 & 175.16$\pm$4.03 & 188.18$\pm$9.31 & 647$\pm$14 & 2403$\pm$40 & 8129$\pm$208 & 41.82$\pm$0.13 & 42.89$\pm$0.11 & 42.92$\pm$0.05 \\
6 & J081652.25+425829.42 & 08 16 52.24 & +42 58 29.41 & 0.23 & 5.35$\pm$0.53 & 162.93$\pm$3.45 & 232.18$\pm$8.37 & 294$\pm$19 & 2421$\pm$37 & 7493$\pm$132 & 41.44$\pm$0.08 & 42.92$\pm$0.07 & 43.07$\pm$0.11 \\
7 & J091955.35+552137.12 & 09 19 55.34 & +55 21 37.12 & 0.12 & 7.66$\pm$0.53 & 173.32$\pm$4.24 & 266.45$\pm$9.98 & 382$\pm$8 & 2777$\pm$46 & 8092$\pm$145 & 41.55$\pm$0.07 & 42.91$\pm$0.14 & 43.09$\pm$0.16 \\
8 & J094636.43+205610.98 & 09 46 36.42 & +20 56 10.89 & 0.28 & 4.86$\pm$1.41 & 169.39$\pm$6.67 & 234.69$\pm$13.46 & 535$\pm$42 & 2575$\pm$66 & 8136$\pm$242 & 41.42$\pm$0.15 & 42.96$\pm$0.14 & 43.10$\pm$0.12 \\
9 & J095048.39+392650.52 & 09 50 48.38 & +39 26 50.46 & 0.21 & 8.31$\pm$0.62 & 182.63$\pm$4.44 & 270.93$\pm$10.64 & 401$\pm$16 & 2811$\pm$47 & 8128$\pm$152 & 41.82$\pm$0.04 & 43.16$\pm$0.10 & 43.34$\pm$0.21 \\
10 & J100420.14+051300.43 & 10 04 20.13 & +05 13 00.39 & 0.16 & 33.90$\pm$6.96 & 153.37$\pm$5.97 & 158.43$\pm$10.94 & 500$\pm$206 & 2455$\pm$53 & 7658$\pm$275 & 42.31$\pm$0.06 & 42.96$\pm$0.17 & 42.98$\pm$0.04 \\
11 & J100642.59+412201.88 & 10 06 42.581 & +41 22 01.95  & 0.15 & 56.54$\pm$2.82 & 125.89$\pm$7.35 & 174.61$\pm$14.91 & 605$\pm$27 & 3018$\pm$107 & 8079$\pm$314 & 42.01$\pm$0.06 & 42.36$\pm$0.03 & 42.50$\pm$0.18 \\
12 & J100726.10+124856.22  & 10 07 26.09 & +12 48 56.18 & 0.24 & 8.93$\pm$0.68 & 91.04$\pm$6.03 & 351.46$\pm$14.54 & 326$\pm$16 & 3417$\pm$153 & 9772$\pm$194 & 42.07$\pm$0.09 & 43.08$\pm$0.16 & 43.67$\pm$0.05 \\
13 & J102531.28+514034.87 & 10 25 31.27 & +51 40 34.86 & 0.04 & 14.26$\pm$1.35 & 117.31$\pm$7.01 & 107.96$\pm$12.17 & 356$\pm$13 & 1255$\pm$60 & 3211$\pm$145 & 40.81$\pm$0.11 & 41.73$\pm$0.08 & 41.69$\pm$0.19 \\
14 & J105143.90+335926.70 & 10 51 43.89 & +33 59 26.68 & 0.17 & 5.62$\pm$0.65 & 196.98$\pm$6.58 & 145.53$\pm$12.51 & 327$\pm$7 & 2498$\pm$53 & 5868$\pm$194 & 41.50$\pm$0.16 & 43.04$\pm$0.06 & 42.91$\pm$0.09 \\
15 & J105151.45-005117.68 & 10 51 51.44 &  -00 51 17.70 & 0.36 & 5.56$\pm$0.48 & 93.89$\pm$4.98 & 196.24$\pm$10.77 & 376$\pm$13 & 2375$\pm$82 & 5772$\pm$125 & 42.52$\pm$0.09 & 43.75$\pm$0.13 & 44.07$\pm$0.06 \\
16 & J105705.41+580437.52 & 10 57 05.40 & +58 04 37.45 & 0.14 & 21.03$\pm$4.96 & 209.19$\pm$10.27 & 46.04$\pm$23.56 & 232$\pm$54 & 2234$\pm$73 & 6086$\pm$1386 & 41.24$\pm$0.17 & 42.24$\pm$0.11 & 41.58$\pm$0.07 \\
17 & J111706.40+441333.31 & 11 17 06.39 & +44 13 33.32 & 0.14 & 7.96$\pm$0.50 & 161.10$\pm$3.58 & 260.39$\pm$8.30 & 408$\pm$7 & 2660$\pm$40 & 7818$\pm$120 & 41.84$\pm$0.10 & 43.15$\pm$0.03 & 43.36$\pm$0.19 \\
18 & J111830.29+402554.01 & 11 18 30.28 & +40 25 54.06 & 0.15 & 11.87$\pm$1.31 & 141.78$\pm$3.03 & 144.75$\pm$7.20 & 425$\pm$36 & 1467$\pm$22 & 5336$\pm$144 & 41.90$\pm$0.02 & 42.97$\pm$0.05 & 42.98$\pm$0.09 \\
19 & J111908.67+211917.9 & 11 19 08.67 & +21 19 17.98 & 0.18 & 7.17$\pm$1.73 & 160.49$\pm$5.04 & 234.21$\pm$11.14 & 823$\pm$45 & 2248$\pm$47 & 7351$\pm$178 & 42.35$\pm$0.13 & 43.70$\pm$0.15 & 43.87$\pm$0.08 \\
20 & J112439.18+420145.03 & 11 24 39.17 & +42 01 45.01 & 0.22 & 10.19$\pm$0.95 & 165.07$\pm$3.51 & 217.18$\pm$8.79 & 477$\pm$32 & 2393$\pm$39 & 7592$\pm$153 & 42.19$\pm$0.20 & 43.40$\pm$0.13 & 43.52$\pm$0.16 \\
21 & J114043.62+532439.03 & 11 40 43.62 & +53 24 38.96 & 0.53 & 6.82$\pm$0.53 & 149.15$\pm$6.55 & 150.68$\pm$20.98 & 394$\pm$12 & 2953$\pm$80 & 7070$\pm$268 & 41.76$\pm$0.13 & 43.10$\pm$0.09 & 43.11$\pm$0.14 \\
22 & J115324.47+493108.76 & 11 53 24.46 & +49 31 08.83 & 0.33 & 8.85$\pm$0.53 & 213.13$\pm$9.65 & 136.12$\pm$19.00 & 395$\pm$11 & 3280$\pm$72 & 6644$\pm$304 & 41.84$\pm$0.12 & 43.22$\pm$0.08 & 43.03$\pm$0.11 \\
23 & J115341.78+461242.25 & 11 53 41.77 & +46 12 42.17 & 0.02 & 26.41$\pm$2.24 & 113.76$\pm$5.10 & 98.14$\pm$9.91 & 254$\pm$15 & 685$\pm$23 & 2272$\pm$109 & 40.46$\pm$0.15 & 41.09$\pm$0.18 & 41.03$\pm$0.05 \\
24 & J120442.11+275411.82 & 12 04 42.10 & +27 54 11.86 & 0.17 & 8.63$\pm$0.78 & 204.56$\pm$6.06 & 279.16$\pm$14.43 & 396$\pm$11 & 3016$\pm$61 & 9055$\pm$231 & 41.53$\pm$0.16 & 42.90$\pm$0.12 & 43.04$\pm$0.07 \\
25 & J122539.56+245836.39 & 12 25 39.54 & +24 58 36.38 & 0.27 & 17.85$\pm$0.60 & 257.55$\pm$11.05 & 250.24$\pm$23.61 & 394$\pm$6 & 4334$\pm$99 & 9650$\pm$345 & 42.11$\pm$0.07 & 43.26$\pm$0.07 & 43.25$\pm$0.04 \\
26 & J123203.63+200929.55 & 12 32 03.62 & +20 09 29.49 & 0.06 & 6.45$\pm$0.50 & 138.07$\pm$3.39 & 240.03$\pm$7.49 & 308$\pm$7 & 2024$\pm$36 & 6096$\pm$92 & 40.90$\pm$0.07 & 42.24$\pm$0.15 & 42.47$\pm$0.08 \\
27 & J124635.25+022208.75 & 12 46 35.25 & +02 22 08.78 & 0.05 & 26.33$\pm$2.97 & 44.24$\pm$2.23 & 145.19$\pm$6.56 & 401$\pm$14 & 355$\pm$15 & 1821$\pm$40 & 40.98$\pm$0.11 & 41.21$\pm$0.13 & 41.72$\pm$0.07 \\
28 & J130947.00+081948.24 & 13 09 47.00 & +08 19 48.21 & 0.15 & 7.62$\pm$0.61 & 186.64$\pm$4.79 & 270.93$\pm$11.37 & 370$\pm$12 & 2823$\pm$49 & 8314$\pm$169 & 41.89$\pm$0.09 & 43.28$\pm$0.10 & 43.44$\pm$0.17 \\
29 & J131217.75+351521.06 & 13 12 17.75 & +35 15 21.08 & 0.18 & 10.33$\pm$0.96 & 142.27$\pm$6.42 & 230.98$\pm$13.90 & 527$\pm$20 & 2858$\pm$81 & 7335$\pm$192 & 42.07$\pm$0.02 & 43.21$\pm$0.07 & 43.42$\pm$0.04 \\
30 & J134356.75+253847.69 & 13 43 56.74 & +25 38 47.63 & 0.09 & 4.59$\pm$0.40 & 205.43$\pm$3.40 & 126.26$\pm$7.34 & 405$\pm$8 & 2554$\pm$27 & 6542$\pm$161 & 40.71$\pm$0.05 & 42.36$\pm$0.19 & 42.15$\pm$0.13 \\
31 & J135550.21+204614.54 & 13 55 50.20 & +20 46 14.51 & 0.2 & 42.38$\pm$6.49 & 160.89$\pm$12.53 & 155.83$\pm$10.85 & 1093$\pm$136 & 2896$\pm$113 & 8732$\pm$291 & 42.79$\pm$0.19 & 43.37$\pm$0.17 & 43.35$\pm$0.21 \\
32 & J135632.80+210352.40 & 13 56 32.79 & +21 03 52.36 & 0.3 & 7.01$\pm$0.78 & 228.43$\pm$11.01 & 271.19$\pm$23.49 & 397$\pm$26 & 3712$\pm$95 & 8423$\pm$281 & 41.87$\pm$0.10 & 43.39$\pm$0.05 & 43.46$\pm$0.20 \\
33 & J140251.20+263117.59 & 14 02 51.19 & +26 31 17.55 & 0.19 & 2.93$\pm$1.76 & 42.07$\pm$4.27 & 472.10$\pm$5.68 & 499$\pm$31 & 1175$\pm$79 & 6756$\pm$52 & 41.25$\pm$0.19 & 42.40$\pm$0.15 & 43.45$\pm$0.06 \\
34 & J140438.80+432707.43 & 14 04 38.79 & +43 27 07.45 & 0.32 & 8.05$\pm$2.94 & 158.98$\pm$5.82 & 144.36$\pm$4.74 & 1076$\pm$163 & 2212$\pm$33 & 7627$\pm$129 & 42.59$\pm$0.03 & 43.89$\pm$0.10 & 43.85$\pm$0.05 \\
35 & J141556.85+052029.56 & 14 15 56.83 & +05 20 29.58 & 0.13 & 6.23$\pm$1.62 & 260.59$\pm$12.15 & 128.13$\pm$31.90 & 266$\pm$20 & 2745$\pm$95 & 11043$\pm$1628 & 40.76$\pm$0.04 & 42.38$\pm$0.13 & 42.07$\pm$0.17 \\
36 & J141700.82+445606.36 & 14 17 00.82 & +44 56 06.33 & 0.11 & 0.78$\pm$0.74 & 158.41$\pm$3.70 & 129.75$\pm$7.36 & 450$\pm$89 & 1747$\pm$29 & 5119$\pm$133 & 40.33$\pm$0.06 & 42.64$\pm$0.17 & 42.56$\pm$0.16 \\
37 & J142748.29+050222.04 & 14 27 48.28 & +05 02 22.04 & 0.11 & 8.33$\pm$1.90 & 96.09$\pm$5.07 & 167.28$\pm$6.03 & 283$\pm$19 & 704$\pm$19 & 2724$\pm$54 & 41.39$\pm$0.04 & 42.45$\pm$0.04 & 42.70$\pm$0.08 \\
38 & J142943.07+474726.20 & 14 29 43.07 & +47 47 26.21 & 0.22 & 8.46$\pm$0.94 & 179.94$\pm$4.42 & 197.19$\pm$10.62 & 444$\pm$22 & 2286$\pm$42 & 7939$\pm$225 & 41.85$\pm$0.16 & 43.17$\pm$0.12 & 43.21$\pm$0.10 \\
39 & J144645.94+403505.76 & 14 46 45.93 & +40 35 05.79 & 0.27 & 11.88$\pm$1.40 & 173.28$\pm$10.38 & 334.82$\pm$23.39 & 394$\pm$6 & 2890$\pm$121 & 9216$\pm$354 & 41.93$\pm$0.08 & 43.09$\pm$0.02 & 43.38$\pm$0.19 \\
40 & J144825.10+355946.65 & 14 48 25.09 & +35 59 46.70 & 0.11 & 9.54$\pm$0.95 & 163.37$\pm$3.98 & 102.99$\pm$8.80 & 365$\pm$24 & 1668$\pm$31 & 5452$\pm$232 & 41.18$\pm$0.18 & 42.42$\pm$0.13 & 42.22$\pm$0.05 \\
41 & J145108.76+270926.92 & 14 51 08.76 & +27 09 26.96 & 0.06 & 9.63$\pm$1.00 & 113.94$\pm$2.79 & 129.02$\pm$5.95 & 268$\pm$9 & 752$\pm$12 & 2848$\pm$67 & 41.31$\pm$0.10 & 42.38$\pm$0.10 & 42.43$\pm$0.12 \\
42 & J145608.65+275008.75 & 14 56 08.65 & +27 50 08.71 & 0.25 & 5.20$\pm$1.51 & 190.49$\pm$3.67 & 122.93$\pm$7.37 & 562$\pm$123 & 2014$\pm$23 & 7353$\pm$244 & 41.57$\pm$0.09 & 43.13$\pm$0.14 & 42.94$\pm$0.15 \\
43 & J152114.26+222743.87 & 15 21 14.25 & +22 27 43.82 & 0.14 & 45.73$\pm$8.31 & 166.73$\pm$4.46 & 158.63$\pm$9.01 & 1173$\pm$209 & 2895$\pm$54 & 8681$\pm$234 & 42.36$\pm$0.19 & 42.93$\pm$0.18 & 42.90$\pm$0.06 \\
44 & J153552.40+575409.50 & 15 35 52.40 & +57 54 09.51 & 0.03 & 21.86$\pm$0.55 & 223.50$\pm$5.37 & 135.34$\pm$12.28 & 385$\pm$5 & 3287$\pm$49 & 8417$\pm$329 & 41.02$\pm$0.07 & 42.03$\pm$0.05 & 41.81$\pm$0.21 \\
45 & J154307.78+193751.75 & 15 43 07.77 & +19 37 51.74 & 0.23 & 7.50$\pm$0.83 & 173.80$\pm$4.39 & 246.19$\pm$10.61 & 437$\pm$23 & 2461$\pm$46 & 8015$\pm$175 & 41.74$\pm$0.19 & 43.11$\pm$0.16 & 43.26$\pm$0.04 \\
46 & J154743.54+205216.66 & 15 47 43.53 & +20 52 16.61 & 0.26 & 9.13$\pm$0.65 & 173.73$\pm$5.99 & 283.73$\pm$13.76 & 425$\pm$7 & 3052$\pm$67 & 8125$\pm$176 & 42.32$\pm$0.17 & 43.60$\pm$0.15 & 43.81$\pm$0.16 \\
47 & J155444.58+082221.45 & 15 54 44.57 & +08 22 21.41 & 0.12 & 34.20$\pm$2.47 & 140.96$\pm$3.29 & 99.52$\pm$6.22 & 535$\pm$36 & 1737$\pm$30 & 5665$\pm$173 & 42.17$\pm$0.13 & 42.78$\pm$0.13 & 42.63$\pm$0.22 \\
48 & J155936.14+544203.77 & 15 59 36.14 & +54 42 03.80 & 0.31 & 7.34$\pm$0.66 & 93.95$\pm$3.77 & 171.59$\pm$8.11 & 265$\pm$6 & 1226$\pm$35 & 3445$\pm$74 & 41.42$\pm$0.06 & 42.53$\pm$0.11 & 42.79$\pm$0.15 \\
49 & J161413.20+260416.20 & 16 14 13.20 & +26 04 16.22 & 0.13 & 26.45$\pm$0.69 & 160.16$\pm$3.84 & 125.52$\pm$8.54 & 351$\pm$6 & 2130$\pm$36 & 6270$\pm$201 & 42.35$\pm$0.11 & 43.13$\pm$0.14 & 43.02$\pm$0.10 \\
50 & J163459.83+204936.09 & 16 34 59.82 & +20 49 36.12 & 0.13 & 14.70$\pm$6.96 & 154.08$\pm$24.00 & 158.69$\pm$34.9 & 820$\pm$57 & 2534$\pm$211 & 6770$\pm$654 & 41.27$\pm$0.02 & 42.29$\pm$0.10 & 42.30$\pm$0.11 \\
51 & J171207.44+584754.51 & 17 12 07.43 & +58 47 54.45 & 0.27 & 33.02$\pm$1.93 & 158.89$\pm$4.92 & 122.47$\pm$9.11 & 586$\pm$24 & 1752$\pm$39 & 5768$\pm$211 & 42.20$\pm$0.18 & 42.88$\pm$0.10 & 42.76$\pm$0.18 \\
52 & J223607.68+134355.3 & 22 36 07.68 & +13 43 55.36 & 0.33 & 7.56$\pm$0.72 & 123.60$\pm$3.27 & 150.22$\pm$6.95 & 303$\pm$11 & 1353$\pm$28 & 4836$\pm$118 & 42.23$\pm$0.01 & 43.45$\pm$0.19 & 43.53$\pm$0.19 \\
53 & J235156.13-010913.35 & 23 51 56.12 & -01 09 13.31 & 0.17 & 11.34$\pm$0.58 & 206.52$\pm$5.92 & 273.71$\pm$12.74 & 406$\pm$10 & 3793$\pm$70 & 9393$\pm$181 & 41.95$\pm$0.15 & 43.21$\pm$0.09 & 43.33$\pm$0.06 \\

\hline
\end{tabular}
\end{table*}

\begin{table*}
\caption{Parameters used in the present study. Column 2: continuum luminosity at 5100 {\AA}; columns 3 and 4: MIR luminosities adopted from Lakicevic et al. (2017) at 6 ${\upmu \text{m}}$ and 12 ${\upmu \text{m}}$, respectively; columns 5 and 6: X-ray luminosity and photon index in 2 - 10 keV energy band; column 7: soft X-ray luminosity adopted from Anderson et al. (2007); columns 8 and 9: FWHMs of broad components of H$\beta$ and H$\alpha$, respectively; column 10: mass of the black hole; column 11: Eddington ratio; column 12: broad Balmer decrement; column 13: references for L$_{2-10\, \text{keV}}$ and $\Gamma$ where (1) Krawczyk et al. (2013); (2) Bianchi et al. (2009); (3) Ricci et al. (2017); (4) Piconcelli et al. (2005); (5) Leighly (1999); (6) Salvato et al. (2018); (7) Vaughan et al. (1999); (8) Young et al. (2009); (9) Zhou \& Zhang (2010); (10) Ruiz et al. (2021).}
\label{tab:T3}
\setlength\tabcolsep{3pt}
\setlength{\cmidrulekern}{2em}
\scriptsize
\begin{tabular}{|l|c|c|c|c|c|c|c|c|c|c|c|c|}
\hline
ID & log($\uplambda \text{L}_{5100}\text{\AA}$) & log(L$_{6\,\upmu \text{m}}$) & log(L$_{12\,\upmu \text{m}}$) & log(L$_{2-10\,\text{keV}}$) & $\Gamma$ & log(L$_{0.1-2.4\,\text{keV}}$) & H$\beta ^\text{B}$ FWHM & H$\alpha ^\text{B}$ FWHM & log(M$_{\text{BH}}$) & $\lambda_{\text{Edd}}$ & log (H$\alpha^{B}/ H\beta^{B}$) & Ref.\\
 & (erg s$^{-1}$) & (erg s$^{-1}$) & (erg s$^{-1}$) & (erg s$^{-1}$) &  & (erg s$^{-1}$) & (km s$^{-1}$) & (km s$^{-1}$) & ($\text{M}_{\odot}$) &  &  &\\
 (1) & (2) & (3) & (4) & (5) & (6) & (7) & (8) & (9) & (10) & (11) & (12) & (13)\\
 \\
 \hline
1 & 44.67$\pm$0.04 & 44.87 & 44.91 & 43.73$\pm$0.17 & --- & --- & 5808$\pm$550 & 5914$\pm$499 & 8.68 & 0.28$\pm$0.02 & 0.55$\pm$0.01 & 1 \\
2 & 44.54$\pm$0.05 & 45.00 & 45.30 & 43.82$\pm$0.08 & 1.8$\pm$0.5 & 44.53 & 4442$\pm$559 & 4539$\pm$427 & 8.36 & 0.49$\pm$0.01 & 0.38$\pm$0.03 & 2 \\
3 & 43.62$\pm$0.06 & 43.86 & 43.76 & 43.59$\pm$0.06 & 1.58$\pm$0.5 & 45.76 & 8263$\pm$518 & 8621$\pm$359 & 8.25 & 0.07$\pm$0.04 & 0.56$\pm$0.01 & 2 \\
4 & 44.06$\pm$0.05 & 44.27 & 44.56 & 43.04$\pm$0.07 & --- & --- & 8162$\pm$238 & 8264$\pm$358 & 8.55 & 0.09$\pm$0.05 & 0.61$\pm$0.03 & 2 \\
5 & 44.28$\pm$0.08 & 44.67 & 44.66 & 43.77$\pm$0.17 & --- & 44.7 & 11858$\pm$505 & 12187$\pm$451 & 9.02 & 0.09$\pm$0.05 & 0.48$\pm$0.02 & 1 \\
6 & 44.59$\pm$0.09 & 44.56 & 44.50 & 43.38$\pm$0.05 & --- & --- & 6463$\pm$223 & 6642$\pm$390 & 8.71 & 0.25$\pm$0.04 & 0.51$\pm$0.04 & 3 \\
7 & 44.51$\pm$0.10 & 44.38 & 44.40 & 43.87$\pm$0.17 & --- & 44.22 & 7289$\pm$365 & 7350$\pm$438 & 8.77 & 0.20$\pm$0.04 & 0.55$\pm$0.03 & 1 \\
8 & 44.58$\pm$0.08 & 45.18 & 45.22 & 44.70$\pm$0.17 & --- & --- & 8329$\pm$210 & 8443$\pm$397 & 8.93 & 0.19$\pm$0.02 & 0.55$\pm$0.02 & 1 \\
9 & 44.71$\pm$0.06 & 44.99 & 44.93 & 44.33$\pm$0.09 & 1.6$\pm$0.2 & 44.84 & 7149$\pm$560 & 7429$\pm$469 & 8.89 & 0.21$\pm$0.01 & 0.54$\pm$0.03 & 2 \\
10 & 44.54$\pm$0.09 & 44.72 & 44.61 & 43.71$\pm$0.07 & --- & 43.49 & 4145$\pm$330 & 5360$\pm$547 & 8.30 & 0.52$\pm$0.04 & 0.30$\pm$0.01 & 4 \\
11 & 44.23$\pm$0.08 & 44.44 & 44.70 & 43.39$\pm$0.17 & --- & --- & 10437$\pm$485 & 10559$\pm$458 & 8.88 & 0.05$\pm$0.04 & 0.70$\pm$0.03 & 1 \\
12 & 45.19$\pm$0.07 & 45.08 & 45.27 & 44.75$\pm$0.09 & 1.57$\pm$0.16 & --- & 7817$\pm$530 & 7886$\pm$425 & 9.30 & 0.25$\pm$0.05 & 0.73$\pm$0.02 & 2 \\
13 & 43.64$\pm$0.06 & 43.48 & 43.49 & 43.53$\pm$0.07 & 2.12$\pm$0.1 & 44 & 4109$\pm$385 & 4479$\pm$397 & 7.66 & 0.30$\pm$0.02 & 0.39$\pm$0.03 & 5 \\
14 & 44.56$\pm$0.07 & 44.42 & 44.43 & 44.01$\pm$0.08 & 1.5$\pm$0.2 & 44.65 & 6806$\pm$543 & 7032$\pm$364 & 8.74 & 0.21$\pm$0.06 & 0.52$\pm$0.03 & 2 \\
15 & 45.59$\pm$0.05 & 45.66 & 45.73 & 45.51$\pm$0.07 & --- & 45.08 & 8280$\pm$379 & 8425$\pm$394 & 9.63 & 0.20$\pm$0.04 & 0.47$\pm$0.04 & 1 \\
16 & 43.68$\pm$0.08 & 43.95 & 44.08 & 43.39$\pm$0.19 & --- & 43.94 & 7061$\pm$276 & 7319$\pm$417 & 8.15 & 0.09$\pm$0.02 & 0.51$\pm$0.02 & 6 \\
17 & 44.78$\pm$0.05 & 44.95 & 44.99 & 44.09$\pm$0.08 & 1.58$\pm$0.05 & 43.3 & 7235$\pm$224 & 7648$\pm$422 & 8.96 & 0.21$\pm$0.03 & 0.64$\pm$0.03 & 2 \\
18 & 44.72$\pm$0.10 & 44.61 & 44.57 & 43.89$\pm$0.09 & 2.14$\pm$0.18 & 44.77 & 6159$\pm$209 & 6587$\pm$380 & 8.77 & 0.29$\pm$0.04 & 0.44$\pm$0.03 & 2 \\
19 & 45.28$\pm$0.09 & 45.33 & 45.21 & 44.46$\pm$0.09 & 1.54$\pm$0.06 & --- & 9370$\pm$340 & 9695$\pm$514 & 9.52 & 0.18$\pm$0.04 & 0.46$\pm$0.01 & 2 \\
20 & 44.96$\pm$0.10 & 44.76 & 44.52 & 44.20$\pm$0.19 & --- & 44.41 & 6468$\pm$405 & 6741$\pm$498 & 8.98 & 0.28$\pm$0.03 & 0.49$\pm$0.02 & 6 \\
21 & 44.77$\pm$0.04 & 45.14 & 45.16 & 44.33$\pm$0.17 & --- & --- & 8949$\pm$307 & 9337$\pm$451 & 9.13 & 0.13$\pm$0.02 & 0.60$\pm$0.02 & 1 \\
22 & 44.68$\pm$0.07 & 45.09 & --- & 45.30$\pm$0.04 & --- & 45.43 & 6624$\pm$285 & 7017$\pm$409 & 8.80 & 0.22$\pm$0.02 & 0.63$\pm$0.02 & 3 \\
23 & 42.93$\pm$0.11 & 42.81 & 42.87 & 42.38$\pm$0.05 & 2.01$\pm$0.17 & 42.63 & 4235$\pm$462 & 5943$\pm$353 & 7.21 & 0.17$\pm$0.04 & 0.49$\pm$0.02 & 7 \\
24 & 44.31$\pm$0.08 & 44.70 & 44.87 & 44.41$\pm$0.07 & 1.56$\pm$0.11 & --- & 8162$\pm$315 & 8548$\pm$413 & 8.73 & 0.12$\pm$0.06 & 0.53$\pm$0.01 & 2 \\
25 & 44.69$\pm$0.10 & 44.26 & --- & 43.87$\pm$0.17 & --- & --- & 9893$\pm$447 & 10287$\pm$469 & 9.16 & 0.08$\pm$0.03 & 0.68$\pm$0.03 & 1 \\
26 & 43.98$\pm$0.07 & 44.12 & 44.11 & 43.46$\pm$0.04 & 1.83$\pm$0.05 & --- & 8184$\pm$254 & 8283$\pm$415 & 8.49 & 0.10$\pm$0.04 & 0.49$\pm$0.01 & 2 \\
27 & 43.60$\pm$0.04 & 43.60 & 43.80 & 43.06$\pm$0.03 & 2.31$\pm$0.04 & 44.05 & 3485$\pm$532 & 3788$\pm$529 & 7.49 & 0.41$\pm$0.03 & 0.34$\pm$0.03 & 2 \\
28 & 44.77$\pm$0.08 & 44.72 & 44.78 & 44.18$\pm$0.09 & 1.51$\pm$0.06 & --- & 7771$\pm$325 & 8099$\pm$423 & 9.00 & 0.19$\pm$0.02 & 0.52$\pm$0.02 & 2 \\
29 & 44.98$\pm$0.07 & 44.96 & 45.13 & 43.82$\pm$0.08 & 1.7$\pm$0.1 & --- & 8947$\pm$533 & 9295$\pm$418 & 9.27 & 0.16$\pm$0.01 & 0.54$\pm$0.01 & 4 \\
30 & 43.94$\pm$0.06 & 43.95 & 44.00 & 44.14$\pm$0.09 & --- & --- & 6163$\pm$281 & 6290$\pm$375 & 8.22 & 0.15$\pm$0.03 & 0.48$\pm$0.03 & 6 \\
31 & 44.92$\pm$0.06 & 45.29 & 45.39 & 44.77$\pm$0.22 & --- & --- & 6229$\pm$364 & 6294$\pm$379 & 8.92 & 0.29$\pm$0.06 & 0.67$\pm$0.02 & 6 \\
32 & 44.86$\pm$0.09 & 44.91 & 45.00 & 44.50$\pm$0.17 & --- & --- & 8931$\pm$498 & 9258$\pm$551 & 9.19 & 0.15$\pm$0.05 & 0.63$\pm$0.03 & 1 \\
33 & 44.53$\pm$0.05 & 44.78 & 44.64 & 43.94$\pm$0.08 & 1.52$\pm$0.15 & --- & 6666$\pm$411 & 6696$\pm$513 & 8.70 & 0.21$\pm$0.04 & 0.70$\pm$0.03 & 8 \\
34 & 45.55$\pm$0.05 & 45.85 & 45.96 & 44.42$\pm$0.17 & --- & --- & 7769$\pm$309 & 6982$\pm$395 & 9.55 & 0.32$\pm$0.04 & 0.56$\pm$0.03 & 1 \\
35 & 43.98$\pm$0.08 & 43.63 & 43.80 & 43.26$\pm$0.18 & --- & 43.94 & 9759$\pm$206 & 10188$\pm$476 & 8.65 & 0.06$\pm$0.03 & 0.56$\pm$0.01 & 6 \\
36 & 44.27$\pm$0.06 & 44.41 & 44.43 & 43.55$\pm$0.05 & 1.77$\pm$0.09 & 44.33 & 6941$\pm$380 & 7055$\pm$365 & 8.55 & 0.16$\pm$0.04 & 0.55$\pm$0.04 & 2 \\
37 & 44.35$\pm$0.05 & 43.79 & 43.68 & 43.65$\pm$0.17 & --- & 43.66 & 5457$\pm$309 & 5760$\pm$379 & 8.40 & 0.28$\pm$0.02 & 0.44$\pm$0.02 & 1 \\
38 & 44.54$\pm$0.09 & 44.72 & 44.78 & 44.19$\pm$0.10 & 1.76$\pm$0.14 & 44.69 & 6958$\pm$304 & 7254$\pm$516 & 8.74 & 0.20$\pm$0.04 & 0.43$\pm$0.02 & 2 \\
39 & 44.69$\pm$0.07 & 45.22 & 45.29 & 44.07$\pm$0.10 & 1.88$\pm$0.11 & 45.02 & 9894$\pm$271 & 10316$\pm$460 & 9.16 & 0.11$\pm$0.05 & 0.58$\pm$0.02 & 2 \\
40 & 44.12$\pm$0.08 & 44.13 & 44.39 & 44.10$\pm$0.14 & --- & 44.15 & 7616$\pm$297 & 7754$\pm$530 & 8.53 & 0.11$\pm$0.01 & 0.71$\pm$0.03 & 6 \\
41 & 44.28$\pm$0.05 & 44.01 & 44.11 & 43.32$\pm$0.04 & 2.11$\pm$0.05 & --- & 4724$\pm$346 & 4995$\pm$502 & 8.23 & 0.36$\pm$0.05 & 0.49$\pm$0.02 & 2 \\
42 & 44.73$\pm$0.10 & 44.74 & 44.92 & 44.48$\pm$0.16 & --- & --- & 7260$\pm$233 & 7660$\pm$536 & 8.91 & 0.21$\pm$0.04 & 0.48$\pm$0.02 & 6 \\
43 & 44.42$\pm$0.06 & 44.73 & 44.66 & 44.50$\pm$0.15 & --- & --- & 9865$\pm$458 & 10168$\pm$521 & 8.96 & 0.09$\pm$0.05 & 0.74$\pm$0.04 & 6 \\
44 & 43.60$\pm$0.06 & 43.52 & 43.63 & 43.54$\pm$0.09 & 1.91 & 43.46 & 7816$\pm$382 & 8021$\pm$418 & 8.19 & 0.08$\pm$0.02 & 0.47$\pm$0.02 & 9 \\
45 & 44.55$\pm$0.05 & 45.35 & 45.30 & 43.98$\pm$0.17 & --- & --- & 7258$\pm$477 & 7633$\pm$529 & 8.79 & 0.20$\pm$0.03 & 0.61$\pm$0.03 & 1 \\
46 & 45.21$\pm$0.07 & 45.05 & 45.03 & 45.16$\pm$0.09 & --- & --- & 7905$\pm$258 & 8149$\pm$362 & 9.32 & 0.24$\pm$0.03 & 0.60$\pm$0.01 & 3 \\
47 & 44.53$\pm$0.08 & 44.31 & 44.29 & 43.78$\pm$0.17 & --- & --- & 5651$\pm$597 & 5746$\pm$511 & 8.55 & 0.27$\pm$0.06 & 0.44$\pm$0.02 & 1 \\
48 & 44.40$\pm$0.04 & 44.55 & 44.67 & 43.68$\pm$0.17 & --- & --- & 4749$\pm$465 & 5135$\pm$422 & 8.31 & 0.37$\pm$0.06 & 0.45$\pm$0.03 & 1 \\
49 & 44.76$\pm$0.07 & 44.62 & 44.68 & 44.36$\pm$0.19 & --- & 44.79 & 7516$\pm$349 & 7594$\pm$470 & 8.97 & 0.19$\pm$0.06 & 0.53$\pm$0.04 & 6 \\
50 & 43.98$\pm$0.04 & 44.74 & 44.71 & 43.61$\pm$0.17 & --- & --- & 7348$\pm$268 & 7445$\pm$437 & 8.40 & 0.12$\pm$0.02 & 0.80$\pm$0.03 & 1 \\
51 & 44.56$\pm$0.05 & 44.53 & 44.64 & 44.01$\pm$0.06 & --- & 44.46 & 6960$\pm$390 & 7231$\pm$487 & 8.76 & 0.20$\pm$0.06 & 0.44$\pm$0.01 & 10 \\
52 & 45.29$\pm$0.10 & 45.24 & 45.34 & 44.23$\pm$0.07 & --- & 44.87 & 5000$\pm$360 & 5164$\pm$518 & 8.98 & 0.26$\pm$0.02 & 0.35$\pm$0.02 & 2 \\
53 & 44.75$\pm$0.04 & 44.94 & 44.90 & 44.89$\pm$0.10 & --- & 45.12 & 9219$\pm$293 & 9512$\pm$537 & 9.14 & 0.14$\pm$0.04 & 0.63$\pm$0.02 & 6 \\

\hline
\end{tabular}

\end{table*}

\begin{landscape}
\begin{table}
\caption{Correlations for optical, MIR, and X-ray parameters.}
\label{tab:landscape}
\setlength\tabcolsep{3pt}
\setlength{\cmidrulekern}{0.5em}
    \centering 

\scriptsize
\begin{tabular}{lccccccccccccccccccccc}
\hline
\\
 & & L$_{5100\text{\AA}}$ & L$_{\,6\upmu\text{m}}$ & L$_{12\,\upmu \text{m}}$ & L$_{2-10\,\text{keV}}$ & L$_{0.1-2.4\,\text{keV}}$ & $\Gamma$ & L H$\beta ^N$ & L H$\beta ^\text{B}$ & EW H$\beta ^\text{N}$ & EW H$\beta ^\text{B}$ & FWHM H$\beta ^\text{N}$ & FWHM H$\beta ^\text{B}$ & L H$\alpha ^N$ & L H$\alpha ^\text{B}$ & EW H$\alpha ^\text{N}$ & EW H$\alpha ^\text{B}$ & FWHM H$\alpha ^\text{N}$ & FWHM H$\alpha ^\text{B}$ & Log(M$_{\text{BH}}$) & $\uplambda_{\text{Edd}}$\\
 \\
 \hline
\\
L$_{5100\,\text{\AA}}$ & $\rho$ & 1 & 0.81 & 0.79 & 0.75 & 0.52 & -0.52 & 0.76 & 0.95 & -0.12 & 0.35 & 0.19 & 0.21 & 0.78 & 0.94 & -0.22 & 0.37 & 0.2 & 0.15 & 0.86 & 0.31 \\
 & P & --- & 1.027E-13 & 9.91E-13 & 1.11E-10 & 0.005 & 0.01 & 3.46E-11 & 6.72E-28 & 0.38 & 0.02 & 0.16 & 0.13 & 2.65E-12 & 1.96E-27 & 0.11 & 0.005 & 0.15 & 0.27 & 6.23E-17 & 0.002 \\
L$_{6\,\upmu \text{m}}$ & $\rho$ & 0.81 & 1 & 0.81 & 0.74 & 0.68 & -0.49 & 0.67 & 0.76 & 0.01 & 0.31 & 0.37 & 0.22 & 0.64 & 0.84 & -0.26 & 0.43 & 0.4 & 0.19 & 0.75 & 0.27 \\
 & P & 1.027E-13 & --- & 1.027E-13 & 2.03E-10 & 7.34E-05 & 0.021 & 1.08E-07 & 0 & 0.5 & 0.02 & 0.006 & 0.11 & 2.16E-07 & 1.89E-15 & 0.056 & 0.001 & 0.005 & 0.155 & 0 & 0.05 \\
L$_{12\,\upmu \text{m}}$ & $\rho$ & 0.79 & 0.94 & 1 & 0.71 & 0.67 & -0.42 & 0.69 & 0.74 & -0.02 & 0.28 & 0.31 & 0.28 & 0.66 & 0.81 & -0.21 & 0.42 & 0.41 & 0.27 & 0.76 & 0.14 \\
 & P & 9.91E-13 & 0 & --- & 6.03E-09 & 1.24E-07 & 0.058 & 4.57E-08 & 0 & 0.87 & 0.04 & 0.02 & 0.04 & 1.19E-07 & 2.04E-13 & 0.13 & 0.001 & 0.007 & 0.05 & 0 & 0.29 \\
L$_{2-10\,\text{keV}}$ & $\rho$ & 0.75 & 0.74 & 0.71 & 1 & 0.66 & -0.73 & 0.64 & 0.75 & -0.04 & 0.36 & 0.27 & 0.29 & 0.61 & 0.75 & -0.05 & 0.4 & 0.26 & 0.26 & 0.73 & 0.47 \\
 & P & 1.11E-10 & 2.03E-10 & 6.03E-09 & --- & 0.0001 & 0.0001 & 1.61E-07 & 4.46E-11 & 0.76 & 0.006 & 0.04 & 0.03 & 7.88E-07 & 2.17E-09 & 0.66 & 0.002 & 0.03 & 0.05 & 3.91E-10 & 0.001 \\
L$_{0.1-2.4\,\text{keV}}$ & $\rho$ & 0.52 & 0.68 & 0.67 & 0.66 & 1 & -0.3 & 0.36 & 0.54 & 0.25 & -0.56 & 0.05 & 0.34 & 0.43 & 0.53 & -0.21 & 0.34 & 0.07 & 0.29 & 0.63 & -0.09 \\
 & P & 0.005 & 7.34E-05 & 1.24E-07 & 0.0001 & --- & 0.31 & 0.059 & 0.003 & 0.21 & 0.007 & 0.79 & 0.08 & 0.02 & 0.004 & 0.28 & 0.08 & 0.62 & 0.13 & 0.0004 & 0.62 \\
$\Gamma$ & $\rho$ & -0.52 & -0.49 & -0.42 & -0.73 & -0.3 & 1 & -0.45 & -0.59 & 0.2 & -0.48 & -0.21 & -0.47 & -0.35 & -0.63 & 0.6 & -0.66 & -0.18 & -0.47 & -0.57 & 0.41 \\
 & P & 0.01 & 0.021 & 0.058 & 0.0001 & 0.31 & --- & 0.03 & 0.005 & 0.37 & 0.02 & 0.34 & 0.03 & 0.03 & 0.001 & 0.001 & 0.001 & 0.12 & 0.02 & 0.006 & 0.006 \\
L H$\beta ^\text{N}$ & $\rho$ & 0.76 & 0.67 & 0.69 & 0.64 & 0.36 & -0.45 & 1 & 0.72 & 0.44 & 0.27 & 0.42 & 0.31 & 0.92 & 0.76 & 0.16 & 0.25 & 0.41 & 0.23 & 0.67 & 0.19 \\
 & P & 3.46E-11 & 1.08E-07 & 4.57E-08 & 1.61E-07 & 0.059 & 0.03 & --- & 5.29E-09 & 0.001 & 0.04 & 0.002 & 0.03 & 4.22E-22 & 4.14E-11 & 0.23 & 0.06 & 0.003 & 0.1 & 0 & 0.16 \\
L H$\beta ^\text{B}$ & $\rho$ & 0.95 & 0.76 & 0.74 & 0.75 & 0.54 & -0.59 & 0.72 & 1 & -0.09 & 0.61 & 0.21 & 0.29 & 0.76 & 0.98 & -0.27 & 0.5 & 0.32 & 0.22 & 0.81 & 0.26 \\
 & P & 6.72E-28 & 0 & 0 & 4.46E-11 & 0.003 & 0.005 & 5.29E-09 & --- & 0.48 & 3.21E-06 & 0.13 & 0.03 & 9.03E-11 & 2.42E-36 & 0.05 & 0.0001 & 0.09 & 0.15 & 0 & 0.05 \\
EW H$\beta ^\text{N}$ & $\rho$ & -0.12 & 0.01 & -0.02 & -0.04 & -0.25 & 0.2 & 0.44 & -0.09 & 1 & 0.01 & 0.25 & 0.13 & 0.34 & -0.1 & 0.67 & 0.3 & 0.18 & 0.01 & 0.01 & -0.15 \\
 & P & 0.38 & 0.5 & 0.87 & 0.76 & 0.21 & 0.37 & 0.001 & 0.48 & --- & 0.89 & 0.07 & 0.33 & 0.01 & 0.45 & 2.73E-08 & 0.19 & 0.19 & 0.9 & 0.9 & 0.28 \\
EW H$\beta ^\text{B}$ & $\rho$ & 0.35 & 0.31 & 0.28 & 0.36 & -0.56 & -0.48 & 0.27 & 0.61 & 0.01 & 1 & 0.05 & 0.41 & 0.31 & 0.42 & -0.21 & 0.67 & 0.06 & 0.31 & 0.37 & -0.14 \\
 & P & 0.02 & 0.02 & 0.04 & 0.006 & 0.007 & 0.02 & 0.04 & 3.21E-06 & 0.89 & --- & 0.72 & 0.005 & 0.01 & 0.001 & 0.04 & 8.71E-08 & 0.25 & 0.01 & 0.006 & 0.29 \\
FWHM H$\beta ^\text{N}$ & $\rho$ & 0.19 & 0.37 & 0.31 & 0.27 & 0.05 & -0.21 & 0.42 & 0.21 & 0.25 & 0.05 & 1 & 0.14 & 0.41 & 0.2 & 0.21 & 0.07 & 0.89 & 0.11 & 0.25 & 0.07 \\
 & P & 0.16 & 0.006 & 0.02 & 0.04 & 0.79 & 0.34 & 0.002 & 0.13 & 0.07 & 0.72 & --- & 0.32 & 0.001 & 0.13 & 0.15 & 0.61 & 3.76E-06 & 0.428 & 0.06 & 0.58 \\
FWHM H$\beta ^\text{B}$ & $\rho$ & 0.21 & 0.22 & 0.28 & 0.29 & 0.34 & -0.47 & 0.31 & 0.29 & 0.13 & 0.41 & 0.14 & 1 & 0.25 & 0.27 & 0.03 & 0.49 & 0.19 & 0.92 & 0.61 & -0.77 \\
 & P & 0.13 & 0.11 & 0.04 & 0.03 & 0.08 & 0.03 & 0.03 & 0.03 & 0.33 & 0.005 & 0.32 & --- & 0.06 & 0.05 & 0.84 & 0.0002 & 0.25 & 2.50E-35 & 2.36E-06 & 0 \\
L H$\alpha ^\text{N}$ & $\rho$ & 0.78 & 0.64 & 0.66 & 0.61 & 0.43 & -0.35 & 0.92 & 0.76 & 0.34 & 0.31 & 0.41 & 0.25 & 1 & 0.74 & 0.3 & 0.16 & 0.38 & 0.21 & 0.76 & 0.17 \\
 & P & 2.65E-12 & 2.16E-07 & 1.19E-07 & 7.88E-07 & 0.02 & 0.03 & 4.22E-22 & 9.03E-11 & 0.01 & 0.01 & 0.001 & 0.06 & --- & 1.25E-10 & 0.02 & 0.23 & 0.002 & 0.12 & 2.52E-11 & 0.19 \\
L H$\alpha ^\text{B}$ & $\rho$ & 0.94 & 0.84 & 0.81 & 0.75 & 0.53 & -0.63 & 0.76 & 0.98 & -0.1 & 0.42 & 0.2 & 0.27 & 0.74 & 1 & -0.28 & 0.55 & 0.32 & 0.22 & 0.88 & 0.23 \\
 & P & 1.96E-27 & 1.89E-15 & 2.04E-13 & 2.17E-09 & 0.004 & 0.001 & 4.14E-11 & 2.42E-36 & 0.45 & 0.001 & 0.13 & 0.05 & 1.25E-10 & --- & 0.04 & 1.96E-05 & 0.08 & 0.1 & 3.75E-18 & 0.04 \\
EW H$\alpha ^\text{N}$ & $\rho$ & -0.22 & -0.26 & -0.21 & -0.05 & -0.21 & 0.6 & 0.16 & -0.27 & 0.67 & -0.21 & 0.21 & 0.03 & 0.3 & -0.28 & 1 & -0.34 & 0.32 & 0.04 & -0.11 & -0.12 \\
 & P & 0.11 & 0.056 & 0.13 & 0.66 & 0.28 & 0.001 & 0.23 & 0.05 & 2.73E-08 & 0.04 & 0.15 & 0.84 & 0.02 & 0.04 & --- & 0.01 & 0.09 & 0.75 & 0.39 & 0.36 \\
EW H$\alpha ^\text{B}$ & $\rho$ & 0.37 & 0.43 & 0.42 & 0.4 & 0.34 & -0.66 & 0.25 & 0.5 & -0.11 & 0.67 & 0.07 & 0.49 & 0.16 & 0.55 & -0.34 & 1 & 0.07 & 0.5 & 0.51 & -0.22 \\
 & P & 0.005 & 0.001 & 0.001 & 0.002 & 0.08 & 0.001 & 0.06 & 0.0001 & 0.39 & 8.71E-08 & 0.61 & 0.0002 & 0.23 & 1.96E-05 & 0.01 & --- & 0.61 & 0.0001 & 0.0001 & 0.09 \\
FWHM H$\alpha ^\text{N}$ & $\rho$ & 0.2 & 0.4 & 0.41 & 0.26 & 0.07 & -0.18 & 0.41 & 0.32 & 0.3 & 0.06 & 0.89 & 0.19 & 0.38 & 0.32 & 0.32 & 0.07 & 1 & 0.25 & 0.27 & 0.07 \\
 & P & 0.15 & 0.005 & 0.007 & 0.03 & 0.62 & 0.12 & 0.003 & 0.09 & 0.19 & 0.25 & 3.76E-06 & 0.25 & 0.002 & 0.08 & 0.09 & 0.61 & --- & 0.12 & 0.04 & 0.46 \\
FWHM H$\alpha ^\text{B}$ & $\rho$ & 0.15 & 0.19 & 0.27 & 0.26 & 0.29 & -0.47 & 0.23 & 0.22 & 0.18 & 0.31 & 0.11 & 0.92 & 0.21 & 0.22 & 0.04 & 0.5 & 0.25 & 1 & 0.56 & -0.81 \\
 & P & 0.27 & 0.155 & 0.05 & 0.05 & 0.13 & 0.02 & 0.1 & 0.15 & 0.19 & 0.01 & 0.428 & 2.50E-35 & 0.12 & 0.1 & 0.75 & 0.0001 & 0.12 & --- & 1.24E-05 & 0 \\
Log(M$_{\text{BH}}$) & $\rho$ & 0.86 & 0.75 & 0.76 & 0.73 & 0.63 & -0.57 & 0.67 & 0.81 & 0.01 & 0.37 & 0.25 & 0.61 & 0.76 & 0.88 & -0.11 & 0.51 & 0.27 & 0.56 & 1 & -0.65 \\
 & P & 6.23E-17 & 0 & 0 & 3.91E-10 & 0.0004 & 0.006 & 0 & 0 & 0.9 & 0.006 & 0.06 & 2.36E-06 & 2.52E-11 & 3.75E-18 & 0.39 & 0.0001 & 0.04 & 1.24E-05 & --- & 1.75E-06 \\
$\uplambda_{\text{Edd}}$ & $\rho$ & 0.31 & 0.27 & 0.14 & 0.47 & -0.09 & 0.41 & 0.19 & 0.26 & -0.15 & -0.14 & 0.07 & -0.77 & 0.17 & 0.23 & -0.12 & -0.22 & 0.07 & -0.81 & -0.65 & 1 \\
 & P & 0.002 & 0.05 & 0.29 & 0.001 & 0.62 & 0.006 & 0.16 & 0.05 & 0.28 & 0.29 & 0.58 & 0 & 0.19 & 0.04 & 0.36 & 0.09 & 0.46 & 0 & 1.75E-06 & --- \\

\hline
\end{tabular}
\end{table}
\end{landscape}

\section*{Acknowledgements}

We thank the anonymous referee for the elaborate and constructive comments that improved the quality of the paper. K.S. acknowledges the financial support of ISRO under AstroSat data utilization program.

Funding for SDSS-III has been provided by the Alfred P. Sloan Foundation, the Participating Institutions, the National Science Foundation, and the U.S. Department of Energy Office of Science. The SDSS-III web site is http://www.sdss3.org/.

SDSS-III is managed by the Astrophysical Research Consortium for the Participating Institutions of the SDSS-III Collaboration including the University of Arizona, the Brazilian Participation Group, Brookhaven National Laboratory, Carnegie Mellon University, University of Florida, the French Participation Group, the German Participation Group, Harvard University, the Instituto de Astrofisica de Canarias, the Michigan State/Notre Dame/JINA Participation Group, Johns Hopkins University, Lawrence Berkeley National Laboratory, Max Planck Institute for Astrophysics, Max Planck Institute for Extraterrestrial Physics, New Mexico State University, New York University, Ohio State University, Pennsylvania State University, University of Portsmouth, Princeton University, the Spanish Participation Group, University of Tokyo, University of Utah, Vanderbilt University, University of Virginia, University of Washington, and Yale University. 
 
\section*{Data Availability}
Data used in this work can be accessed from https://www.sdss.org/dr12/ and is also available with authors.

\bsp	
\label{lastpage}
\end{document}